\documentclass{article}
\usepackage{color, xcolor, colortbl}
\usepackage{graphicx}
\usepackage{geometry}
\usepackage{amsmath,amssymb,amsthm,amsfonts}
\usepackage{algorithm}
\usepackage{algorithmic}
\usepackage{dcolumn}
\usepackage{epstopdf}
\usepackage{bm}
\usepackage[caption=false]{subfig}
\usepackage{appendix}
\usepackage{multirow}
\usepackage{braket}
\usepackage[english]{babel}
\usepackage{hyperref}
\usepackage[capitalize]{cleveref}
\usepackage[square,numbers]{natbib}
\usepackage[T1]{fontenc}
\usepackage{xpatch}
\usepackage{tikz}
\usepackage{adjustbox}
\usepackage{xspace}
\usepackage[roman]{complexity}
\usepackage{xparse}

\newcommand{\Tr}{\operatorname{Tr}}

\newcommand{\opr}[1]{\ensuremath{\operatorname{#1}}}
\newcommand{\mc}[1]{\mathcal{#1}}

\newcommand{\wt}[1]{\widetilde{#1}}

\newcommand{\abs}[1]{\left\lvert#1\right\rvert}
\newcommand{\norm}[1]{\left\lVert#1\right\rVert}

\newcommand{\ud}{\,\mathrm{d}}
\newcommand{\Or}{\mathcal{O}}

\newcommand{\NN}{\mathbb{N}}
\newcommand{\RR}{\mathbb{R}}
\newcommand{\CC}{\mathbb{C}}

\theoremstyle{plain}

\theoremstyle{plain}

\theoremstyle{plain}

\theoremstyle{plain}
\newtheorem*{lem*}{\protect\lemmaname}
\theoremstyle{plain}

\theoremstyle{plain}

\theoremstyle{definition}

\providecommand{\definitionname}{Definition}
\providecommand{\assumptionname}{Assumption}
\providecommand{\corollaryname}{Corollary}
\providecommand{\lemmaname}{Lemma}
\providecommand{\propositionname}{Proposition}
\providecommand{\remarkname}{Remark}
\providecommand{\examplename}{Example}
\providecommand{\theoremname}{Theorem}
\providecommand{\conjecturename}{Conjecture}

\newcommand{\REV}[1]{\textcolor{red}{#1}}

\NewDocumentCommand{\ketbra}{mG{#1}}{\mathinner{|{#1}\rangle\!\langle{#2}|}}

\usetikzlibrary{quantikz}

\usetikzlibrary{fit}
\tikzset{%
  highlight/.style={rectangle,rounded corners,fill=blue!15,draw,fill opacity=0.3,thick,inner sep=0pt}
}

\renewcommand{\REV}[1]{#1}

\begin{document}

\title{Dissipative Preparation of Many-Body Quantum States: Towards Practical Quantum Advantage}
\author{Lin Lin
\thanks{Department of Mathematics, University of California, Berkeley; Applied Mathematics and Computational Research Division, Lawrence Berkeley National Laboratory; Email: \texttt{linlin@math.berkeley.edu}}}

\maketitle

\begin{abstract}
  While dissipation has traditionally been viewed as an obstacle to quantum coherence, it is increasingly recognized as a powerful computational resource. Dissipative protocols can prepare complex many-body quantum states by leveraging engineered system-environment interactions. This essay focuses on a class of algorithms that utilize algorithmically constructed Lindblad generators, and highlight recent advances enabling the preparation of ground and thermal states for certain non-commuting Hamiltonians with rigorous performance guarantees.  We also propose extensions of these protocols to prepare excited and resonance states, which may offer new pathways toward realizing practical quantum advantage on early fault-tolerant quantum computing platforms.
\end{abstract}

\section{Introduction}

The ability to prepare quantum many-body states is a foundational challenge in quantum computation and simulation across condensed matter physics, quantum chemistry, materials science, and high-energy physics. From approximating the ground state of strongly correlated electron systems to simulating thermal behavior at finite temperatures or accessing excited and resonance states relevant to spectroscopy, efficient quantum state preparation  lies at the heart of  many of the most computationally demanding tasks in quantum science. Yet, despite substantial theoretical and algorithmic progress, the design of scalable and reliable quantum algorithms for preparing such states, particularly in regimes that are challenging for classical methods, remains a central challenge towards realizing practical advantages.

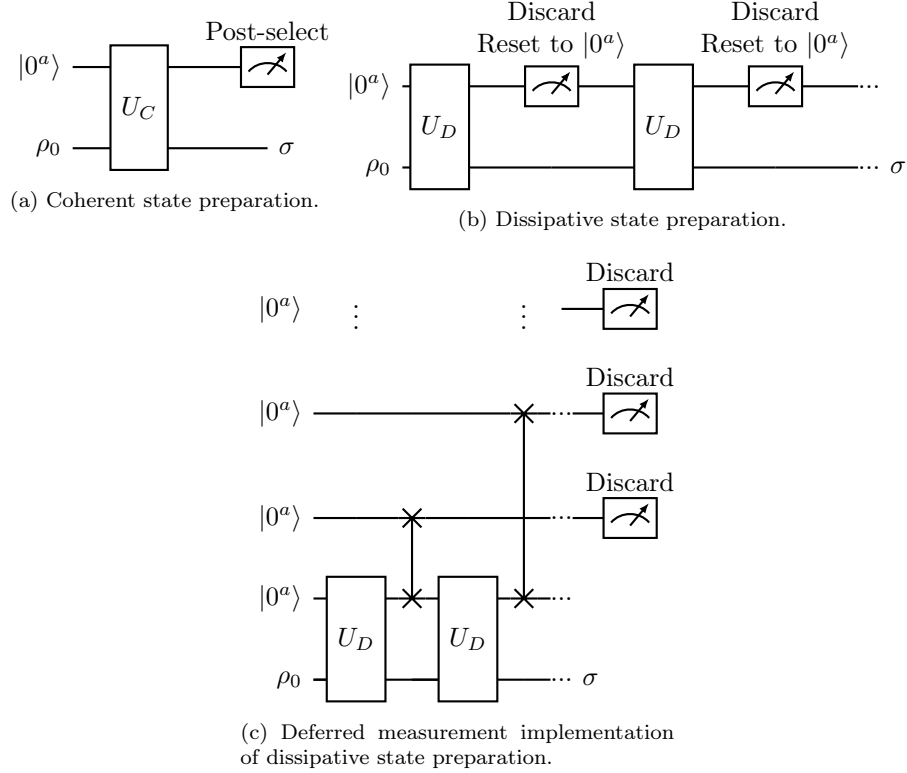
\begin{figure}[H]
\centering
\subfloat[Coherent state preparation.]{
\begin{quantikz}
\lstick{$\ket{0^a}$} & \gate[2]{U_C} & \meter{Post-select}\\
\lstick{$\rho_0$} &  & \qw \rstick{$\sigma$}
\end{quantikz}
}
\subfloat[Dissipative state preparation.]{
\begin{quantikz}[column sep=0.3em]
\lstick{$\ket{0^a}$} & \gate[2]{U_D} & \meter{Discard\\ Reset to $\ket{0^a}$} & \gate[2]{U_D} & \meter{Discard\\Reset to $\ket{0^a}$} & \qw ...\\
\lstick{$\rho_0$} &  & \qw &  & \qw &\qw ... \rstick{$\sigma$}
\end{quantikz}
}

\subfloat[Deferred measurement implementation of dissipative state preparation.]{
\begin{quantikz}[column sep=0.5em]
\lstick{$\ket{0^a}$} &\vdots & &&\vdots &&\meter{Discard}\\
\lstick{$\ket{0^a}$} & \qw & \qw & \qw &  \swap{2} & \qw ... &\meter{Discard}\\
\lstick{$\ket{0^a}$} & \qw & \swap{1} & \qw & \qw & \qw ... &\meter{Discard}\\
\lstick{$\ket{0^a}$} & \gate[2]{U_D} & \targX{} & \gate[2]{U_D} & \targX{} & \qw ...\\
\lstick{$\rho_0$} &  \qw  & \qw &  \qw   & \qw &\qw ... \rstick{$\sigma$}
\end{quantikz}
}
\caption{Schematic representation of (a) coherent state preparation via unitary evolution and post-selection, (b) dissipative state preparation via measurement and trace-over of ancillary degrees of freedom, and (c) conceptually equivalent implementation of dissipative state preparation by introducing more ancilla registers and deferring all measurements to the end. The initial state is $\rho_0$, and the target state is $\sigma$. Both $\rho_0$ and $\sigma$ can be pure or mixed states.}
\label{fig:sketch_coherent_dissipative}
\end{figure}

Most quantum algorithms developed to date, such as quantum phase estimation (QPE)~\cite{NielsenChuang2000} and quantum singular value transformation (QSVT)~\cite{GilyenSuLowEtAl2019}, rely on the paradigm of coherent control. In this framework, a unitary circuit acts on the system register along with one or more ancillary qubits. At the end of the computation, the ancilla is measured, and a post-selection step is carried out based on the measurement outcome. If the measurement does not yield the desired state, the algorithm is restarted and repeated until success (\cref{fig:sketch_coherent_dissipative} (a)).

An alternative approach, known as dissipative state preparation, has also been developed for over two decades, with many applications such as preparation of ground states of stabilizer codes and tensor network states~\cite{KrausBuchlerDiehlEtAl2008,VerstraeteWolfCirac2009,RoyChalkerGornyiEtAl2020,WangSnizhkoRomitoEtAl2023}, many-body states with long-range entanglement \cite{LuLessaKimEtAl2022,foss2023experimental}, and quantum error correction~\cite{BarnesWarren2000,LeghtasKirchmairVlastakisEtAl2013,LieuLiuGorshkov2024}.  
 Dissipative processes can be understood from multiple perspectives. Physically, they correspond to coupling the system to an engineered environment, followed by tracing out the environment to obtain reduced dynamics for the system. There is also a digital viewpoint, which aligns more closely with the circuit model of quantum computation. In this setting, a unitary operation is applied to the combined system and ancilla registers, where the ancilla can be viewed as a digital version of the environment. After measuring the ancilla register, no post-selection is performed, and the algorithm proceeds regardless of the measurement outcome. In other words, the success probability is always $1$ and the process does not require starting from a good initial state. This effectively traces out the ancillas and induces a non-unitary transformation on the system register. By resetting the ancillas and  repeating this process many times, the system can be driven toward the desired target state under appropriate conditions (\cref{fig:sketch_coherent_dissipative} (b)).

Dissipative state preparation is \emph{not} a fundamentally different computational model. By introducing a number of ancilla registers, swapping them in place of mid-circuit measurements, and measuring the ancillas at the end, the dissipative circuit shown in \cref{fig:sketch_coherent_dissipative}(b) can be mapped to a coherent circuit as in \cref{fig:sketch_coherent_dissipative}(a). The key difference is that there is no postselection: the process succeeds regardless of the measurement outcomes of the ancilla registers (\cref{fig:sketch_coherent_dissipative}(c)). 

Despite this conceptual connection, as we will discuss in \cref{sec:lindblad}, the motivation and analysis underlying dissipative preparation differ substantially from those in the coherent framework. In particular, dissipative models naturally lead to non-unitary dynamics and fixed-point problems, which gives rise to very different design principles and convergence analyses.

This essay is not intended as a comprehensive review of the vast and rapidly growing literature on dissipative state preparation or open quantum system dynamics. These fields span a wide array of physical models, mathematical frameworks, and experimental platforms, many of which lie beyond the scope of this work. For a foundational treatment, we refer readers to the standard text~\cite{BreuerPetruccione2002}, as well as to recent comprehensive reviews~\cite{BreuerLainePiiloEtAl2016,DeVegaAlonso2017,WeimerKshetrimayumOrus2021,LandiPolettiSchaller2022}, which survey a range of physical implementations and theoretical strategies.

Our goal, instead, is to highlight recent developments that link the task of quantum state preparation to broader challenges in quantum simulation, and to offer perspectives aimed at a more rigorous understanding of the performance of dissipative protocols. We will focus on a specific class of dissipative dynamics governed by the Lindblad master equation. Rather than restricting attention to exactly solvable models or systems with special, such as commutative or frustration-free structures, we will discuss algorithms applicable to general many-body Hamiltonians that lack such simplifying features, and demonstrate how engineered dissipation can be harnessed in these settings.

The rest of the paper is organized as follows. In \cref{sec:lindblad}, we introduce the Lindblad formalism, discuss methods for simulating Lindblad dynamics on fault-tolerant quantum computers, and analyze state preparation as a fixed-point process with associated cost considerations. In \cref{sec:exam_diss}, we provide a few examples of recent applications of Lindblad dynamics for dissipative state preparation, including protocols for preparing ground states, and Gibbs states, and examine the algorithmic implications of using local versus quasi-local jump operators.
We also extend these ideas to settings such as the preparation of excited states and ground singular vectors and eigenvectors of non-Hermitian matrices. In \cref{sec:experiment}, we discuss how dissipative protocols may be applied to demonstrate practical quantum advantage and explore design considerations for their implementation on early fault-tolerant quantum devices. In \cref{sec:outlook}, we provide an outlook on some future directions in the theory, implementation, and application of dissipative quantum algorithms.

\section{Lindblad dynamics and its quantum simulation}\label{sec:lindblad}
We focus on a specific class of quantum channels governed by Lindblad dynamics, a widely used framework for modeling Markovian open quantum systems. Lindblad dynamics has been employed extensively to capture dissipative and decoherence effects in a variety of settings, including quantum statistical mechanics, quantum optics, circuit-level noise modeling, and quantum error mitigation. In this perspective, we highlight a different role: using Lindblad dynamics as a computational tool for designing new algorithms for quantum state preparation.

\subsection{Lindblad dynamics as an algorithmic tool}\label{sec:lindblad_algorithm}

The Gorini-Kossakowski-Sudarshan-Lindblad (GKSL) master equation, commonly referred to as the Lindblad equation, governs the Markovian open quantum dynamics of a density matrix $\rho$. It takes the canonical form \cite{Lindblad1976,GoriniKossakowskiSudarshan1976}
\begin{equation}\label{eq:lindblad}
    \frac{\ud}{\ud t} \rho = \underbrace{-i[G, \rho]}_{\mathcal{L}_G(\rho)} + \sum_{j=1}^J \underbrace{\left(K_j \rho K_j^\dagger - \frac{1}{2} \{K_j^\dagger K_j, \rho \} \right)}_{\mathcal{L}_{K_j}(\rho)} =: \mathcal{L}(\rho),
\end{equation}
where $G$ is a Hermitian matrix representing the coherent part of the dynamics, and $\{K_j\}_{j=1}^J$ are the so-called jump operators that define the dissipative interactions. The terms $[A, B]$ and $\{A, B\}$ denote the commutator and anticommutator, respectively. The full generator $\mathcal{L}$ is referred to as the Lindbladian, with $\mathcal{L}_G$ and $\mathcal{L}_{K_j}$ denoting its coherent and dissipative components. Given an initial state $\rho(0)$, which may be pure or mixed, the solution to \cref{eq:lindblad} can be formally expressed as
\begin{equation}
    \rho(t) = e^{\mathcal{L} t}[\rho(0)].
\end{equation}

The Lindblad dynamics in \cref{eq:lindblad} can arise in many physical and algorithmic settings. A common physical origin is the reduced dynamics of an open quantum system weakly coupled to a bath. Consider a system-bath Hamiltonian of the form
\begin{equation}\label{eqn:Ht_sysenv}
H(t) = H_S \otimes I_E + I_S \otimes H_E + g A_S \otimes B_E,
\end{equation}
where $H_S$ and $H_E$ are the Hamiltonians of the system and environment, respectively, $A_S$ and $B_E$ are, for simplicity, Hermitian operators characterizing the coupling, and $g$ is the coupling strength. Evolving the total system under unitary dynamics and tracing out the bath yields a reduced dynamics for the system alone. In practice, the environment may have infinitely many degrees of freedom. When the coupling $g$ is sufficiently small, 
a sequence of approximations leads to a master equation of Lindblad form \cite{BreuerPetruccione2002,Lidar2019}. These approximations include the Born approximation (neglecting higher-order interactions), the Markov approximation (neglecting memory effects),  the secular or rotating wave approximation (neglecting interference between widely separated frequencies), together with a weak coupling assumption (\REV{for example, $g$ is smaller than the smallest difference between two Bohr frequencies~\cite{Lidar2019}, which is sometimes referred to as ultra-weak coupling.}). This resulting generator $\mathcal{L}$ is referred to as the \emph{Davies generator} \cite{Davies1974,Davies1976}.

These approximations might suggest that Lindblad dynamics apply only in restricted settings. However, the Lindblad formalism encompasses a much broader class of quantum evolutions.
 As proven in the foundational works~\cite{Lindblad1976,GoriniKossakowskiSudarshan1976}, the generator of \emph{any} quantum dynamical semigroup (QDS), which is a continuous family of quantum channels, must take the Lindblad form in \cref{eq:lindblad}. Moreover, simulating Lindblad dynamics is \BQP-complete~\cite{VerstraeteWolfCirac2009}, meaning that any quantum circuit can be efficiently encoded into a Lindbladian evolution with only polynomial overhead.

In this broader context, Lindblad dynamics can be understood not only as a model of physical open systems of the form \eqref{eqn:Ht_sysenv} in some limiting regimes, but also as a general-purpose algorithmic framework for generating quantum channels. Just as Hamiltonian dynamics are often used in simulation without regard to a specific physical system, Lindblad dynamics can be designed without referencing an actual system-bath interaction. The bath can be viewed as a fictitious and  engineered resource akin to artificial thermostats in classical Monte Carlo or molecular dynamics simulations~\cite{FrenkelSmit2002}.

\subsection{Simulation of Lindblad dynamics on fault-tolerant quantum computers}\label{sec:lindblad_simulate}

How can one simulate Lindblad dynamics of the form \cref{eq:lindblad} on a fault-tolerant quantum computer? For simplicity, let us consider the special case where $G = 0$ and there is a single jump operator $K$. The generalization to multiple jump operators and nonzero $G$ is conceptually straightforward.
Define the Hamiltonian
\begin{equation}
\wt{H} = \begin{pmatrix}
0 & K^\dagger \\
K & 0
\end{pmatrix}.
\end{equation}
Using this Hamiltonian, we construct a quantum channel $\Phi$ as follows:
\begin{equation}\label{eqn:lindblad_hamsim_first}
U_D = e^{-i \wt{H} \sqrt{\Delta t}}, \quad 
\Phi(\rho) = \Tr_a \left[ U_D \left( \ketbra{0}{0} \otimes \rho \right) U_D^\dagger \right],
\end{equation}
where $\Tr_a$ denotes the partial trace over the ancilla subsystem. A Taylor expansion shows that
\begin{equation}
\norm{e^{\mathcal{L} \Delta t}(\rho) - \Phi(\rho)}_1 \leq C (\Delta t)^2,
\end{equation}
for any system density matrix $\rho$, where $\norm{\cdot}_1$ denotes the trace norm. Since quantum channels are contractive with respect to the trace norm, the accumulated simulation error over a total evolution time $T = r \Delta t$ satisfies
\begin{equation}
\norm{e^{\mathcal{L} T}(\rho) - \Phi^r(\rho)}_1 \leq C r (\Delta t)^2 = C T \Delta t.
\end{equation}

This scheme provides a first-order accurate simulation of Lindblad dynamics in $\Delta t$. More importantly, it demonstrates that dissipative processes governed by Lindblad dynamics can be simulated using coherent Hamiltonian evolution on a larger Hilbert space, as illustrated in \cref{fig:sketch_coherent_dissipative}(b), where $U_D$ is realized via Hamiltonian simulation.

Early developments of quantum algorithms for simulating Lindblad equations focused on Trotter expansion methods~\cite{KlieschBarthelGogolinEtAl2011}. More recent approaches go beyond Trotterization and can often be understood through the lens of the Stinespring dilation theorem~\cite{Wolf2012,Watrous2018}, which asserts that any quantum channel can be realized as a unitary evolution on an extended Hilbert space that includes both the system and an ancillary register. From this viewpoint, \cref{eqn:lindblad_hamsim_first} provides the simplest example of a Stinespring dilation for Lindblad dynamics. The ancillary register acts as a fictitious bath with which the system interacts.

High-order methods for simulating Lindblad dynamics on fault-tolerant quantum computers~\cite{CleveWang2017,LiWang2023,DingLiLin2024} can be  understood within the same Stinespring dilation framework. Conceptually, these algorithms design a unitary operator $U_D$ acting on the system and a suitably chosen ancillary register such that
\begin{equation}\label{eqn:stinespring_highorder}
\norm{e^{\mathcal{L} \Delta t}(\rho) - \Phi(\rho)}_1 \leq C_p (\Delta t)^{p+1}, \quad \Phi(\rho) = \Tr_a \left[ U_D \left( \ketbra{0^a}{0^a} \otimes \rho \right) U_D^\dagger \right],
\end{equation}
for any density matrix $\rho$ and arbitrarily large integer $p$. This achieves high-order accuracy with respect to the simulation time step $\Delta t$. Specifically, let $\|\mathcal{L}\|_{\mathrm{be}} := \|G\| + \frac{1}{2}\sum_{j}\|L_j\|^2$, then the query complexity of the algorithm in \cite{LiWang2023} for simulating the Lindblad dynamics \eqref{eq:lindblad} up to time $t$ with precision $\epsilon$ is $\mathcal{O}\left(t\|\mathcal{L}\|_{\mathrm{be}}\log\left(\frac{t\|\mathcal{L}\|_{\mathrm{be}}}{\epsilon}\right)\right)$.


In the following discussion, we focus on the design of \emph{continuous-time} Lindbladian dynamics for efficient state preparation. In doing so, we do not distinguish conceptually between continuous-time Lindblad dynamics and their digital simulation. Thanks to the availability of high-order simulation schemes, an efficient continuous-time dissipative evolution can be directly translated into an efficient quantum algorithm, at least on fault-tolerant quantum computers.

\subsection{Dissipative state preparation as a fixed-point process and cost analysis}

Dissipative state preparation often prepares the target state $\sigma$, which may be pure or mixed, as a unique \emph{fixed point} of a carefully designed dynamics. In the continuous-time setting, this means
\begin{equation}
\partial_t \sigma = \mathcal{L}(\sigma) = 0.
\end{equation}
In a digital implementation using a unitary $U_D$ as in \cref{eqn:stinespring_highorder}, the target state should similarly satisfy an approximate fixed-point property:
\begin{equation}\label{eqn:fixed_point_property}
\Phi(\sigma) = \Tr_a \left[ U_D \left( \ketbra{0^a} \otimes \sigma \right) U_D^\dagger \right] \approx \sigma.
\end{equation}

Recall that in the coherent setting, starting from an initial state $\rho_0$, one applies a unitary circuit $U_C$ and measures the ancilla register. The success probability of observing the ancilla in the all-$0$ state is
\begin{equation}
p_{\text{succ}} = \Tr \left[ (\ketbra{0^a} \otimes I) U_C (\ketbra{0^a} \otimes \rho_0) U_C^\dagger \right],
\end{equation}
and the corresponding post-measurement state in the system register is
\begin{equation}
\sigma \approx (\bra{0^a} \otimes I) U_C (\ketbra{0^a} \otimes \rho_0) U_C^\dagger (\ket{0^a} \otimes I).
\end{equation}
When the success probability $p_{\text{succ}}$ is small, many repetitions may be required. The total cost can be estimated as
\begin{equation}\label{eqn:cost_coherent}
\text{cost}(U_C) \times \text{number of repetitions} \sim \frac{\text{cost}(U_C)}{p_{\text{succ}}}.
\end{equation}
Amplitude amplification can reduce the dependence to $\mathcal{O}(p_{\text{succ}}^{-1/2})$, but it is not possible to achieve further \emph{general} improvement in $p_{\text{succ}}$.

For dissipative algorithms, since the target state is encoded as the fixed point of the dynamics, the time to reach $\sigma$ from an initial state $\rho_0$ is related to the concept of \emph{mixing time}, which can be defined  in terms of the trace distance:
\begin{equation}\label{eqn:mixing_tracedistance}
\tau_{\operatorname{mix}}(\eta) = \min \Big\{t \mid \norm{e^{\mathcal{L}t} (\rho)- \sigma}_1 \leq \eta,\;\; \text{from \emph{any} initial state } \rho \Big\}.
\end{equation}
The discussion of mixing time is often made with respect to a fixed target precision $\eta$. For instance, roughly speaking, for a quantum many-body system on $n$ sites (spins, bosons, fermions, etc), the dynamics is said to exhibit \emph{fast mixing} if $\tau_{\operatorname{mix}} = \Or(\poly(n))$, and \emph{rapid mixing} if $\tau_{\operatorname{mix}} = \Or(\polylog(n))$.

The total cost of dissipative state preparation is then roughly
\begin{equation}\label{eqn:cost_dissipative}
\text{cost}(U_D) \times \text{number of iterations} \sim \text{cost}(U_D) \times \text{mixing time}.
\end{equation}
Although the cost expressions in \cref{eqn:cost_coherent,eqn:cost_dissipative} share a similar form, they differ in several important respects:

\begin{enumerate}

\item \textbf{Dependence on the initial state}: Many coherent state preparation algorithms require the initial state $\rho_0$ to satisfy certain properties. For instance, the problem of ground state preparation for local Hamiltonians is known to be \QMA-hard~\cite{KitaevShenVyalyi2002,AharonovNaveh2002,KempeKitaevRegev2006} in the worst case. However, the sources of complexity differ in coherent versus dissipative settings.
 Many ground state preparation algorithms based on coherent circuits, such as those employing QPE or QSVT, require a good initial state $\rho_0$~\cite{OBrienTarasinskiTerhal2019,GeTuraCirac2019,LinTong2020,DongLinTong2022,LinTong2022,WanBertaCampbell2022}. If $\rho_0$ has zero overlap with the ground state, then  $p_{\text{succ}}=0$. Consequently, a central classical heuristic is whether such an initial state $\rho_0$ can be efficiently constructed using available classical information, and once this assumption is made, whether a quantum algorithm still offers an advantage~\cite{LeeLeeZhaiEtAl2023,Chan2024}.

In contrast, if a dissipative protocol admits a unique fixed point, it can converge from \emph{any} initial state $\rho_0$, including those with zero or exponentially small overlap with the target state. This is reflected in the definition of mixing time in~\eqref{eqn:mixing_tracedistance}, which captures the \emph{maximum} time needed to reach a target distance $\eta$ from any starting point. That being said, for problems that are \NP-hard or \QMA-hard, exponentially long mixing times are to be expected.

\item \textbf{Number of queries to the initial state}: 
In the coherent setting, it is possible, although typically with low probability, for the protocol to succeed in a single run, producing the target state $\sigma$ immediately upon measurement. On the other hand, each unsuccessful outcome, in which the ancilla register is not projected to $\ket{0^a}$, requires restarting the process from the same initial state $\rho_0$. So the expected number of repetitions, as well as the expected number of queries to the initial state $\rho_0$ are $\mathcal{O}(p_{\text{succ}}^{-1})$. The cost to the number of queries to $\rho_0$ can become significant when this initial state preparation step is expensive.

In the dissipative setting, by contrast, this restart overhead can be avoided. Each application of the dissipative map $U_D$ makes an incremental update to the system, gradually steering it toward the target state. But the state in the system register is never discarded in intermediate steps. Each iteration builds upon the state from the previous one, and only a single copy of $\rho_0$ is required. 

\item \textbf{Leveraging fine-grained structure of physical systems}:  The performance of coherent state preparation protocols often depends only on coarse-grained characteristics of the system, such as the spectral gap of the Hamiltonian and the overlap between the initial and target states. Once these parameters are fixed, it often becomes difficult to incorporate more fine-grained structural information about the physical system to further reduce the cost.

This limitation is even more pronounced in the task of Gibbs state preparation, where the target state is given by
\begin{equation}\label{eqn:gibbs}
\sigma_\beta = \frac{e^{-\beta H}}{\mc{Z}_\beta}, \quad \mc{Z}_\beta = \Tr(e^{-\beta H}),
\end{equation}
with $\beta$ the inverse temperature and $\mc{Z}_\beta$ the partition function. For a Hamiltonian $H \in \mathbb{C}^{2^n \times 2^n}$, the cost of coherent-state preparation algorithms \cite{PoulinWocjan2009,ChowdhurySomma2017,VanApeldoornGilyenGriblingEtAl2017,GilyenSuLowEtAl2019,an2023quantum}  includes a prefactor of the form $\Or(\sqrt{2^n/\mc{Z}_\beta})$. While such algorithms can be efficient in the high-temperature regime (i.e., very small $\beta$), this factor often becomes exponentially large even in the moderate temperature regime.

In contrast, dissipative protocols is more analogous to Markov chain Monte Carlo (MCMC) methods in classical computing \cite{LevinPeres2017}, and their performance can be highly problem-specific. While this makes rigorous analysis more challenging and often reliant on empirical evaluation, it also creates opportunities: for certain physically relevant systems, these protocols can dramatically outperform worst-case complexity bounds. Much of the current interest in quantum Gibbs sampling (see \cref{eqn:gibbs_state}) stems from recent developments of Lindblad based algorithms that can bypass this exponentially large prefactor, potentially enabling efficient sampling in physically meaningful temperature regimes.

\item \textbf{Noise resilience and stability of the protocol:}
 In the coherent setting, small coherent errors or gate imperfections typically accumulate linearly in time, with no inherent mechanism for error suppression or cancellation. The dissipative setting does not eliminate the need for quantum error correction. However, it introduces an additional layer of robustness due to its fixed-point structure. Mild implementation errors or gate imperfections can often be modeled as perturbations to the target quantum channel, which typically do not alter the fixed point significantly. As a result, the system can remain close to the target state even when the dissipative dynamics are applied for durations that far exceed the inverse noise rate~\cite{CubittLuciaMichalakisEtAl2015,TrivediFrancoRubioCirac2024,KashyapStyliarisMouradianEtAl2025}.
\end{enumerate}

\section{Dissipative state preparation via Lindblad dynamics}\label{sec:exam_diss}

In this section, we showcase several examples of dissipative preparation of quantum many-body states. These methods do not assume special structures of Hamiltonian (e.g., commuting, frustration-free), and can be applied to a general class of Hamiltonians.

\subsection{Ground state preparation}\label{sec:ground_state}

Recently, a number of protocols~\cite{RoyChalkerGornyiEtAl2020,Cubitt2023,MiMichailidisShabaniEtAl2024,ChenHuangPreskillEtAl2024,DingChenLin2024,li2024dissipative,LambertCirioLinEtAl2024,EderFinzgarBraunEtAl2025,motlagh2024ground,LloydMichailidisMiEtAl2025} have been proposed for dissipative ground state preparation. In this section, we focus on the procedure proposed in~\cite{DingChenLin2024} as a concrete example. In fact, all constructions discussed in the remainder of the paper follow a similar underlying algorithmic structure.

Given a Hamiltonian $H$, a jump operator  in~\cite{DingChenLin2024} takes the form
\begin{equation}\label{eq:jump_operators}
K = \sum_{i, j} \hat{f}(\lambda_i - \lambda_j) \ket{\psi_i} \bra{\psi_i} A \ket{\psi_j} \bra{\psi_j},
\end{equation}
where $\{\lambda_i, \ket{\psi_i}\}$ are the eigenpairs of $H$ ordered \REV{such that $\lambda_0 < \lambda_1 \leq \cdots $}, and $\hat{f}(\omega)$ is a filter function in the frequency domain that vanishes for $\omega > 0$. For a gapped system with a non-degenerate ground state, it suffices to know a lower bound on the spectral gap $\Delta$ and an upper bound on $\norm{H}$ in order to construct $\hat{f}(\omega)$. The resulting jump operator $K$ only allows transitions from higher-energy eigenstates to lower-energy ones. Once the system reaches the ground state $\sigma = \ket{\psi_0} \bra{\psi_0}$, we have $K[\sigma] = 0$, making $\sigma$ a fixed point of the dynamics.

The transition rate from a high-energy state $\ket{\psi_j}$ to a low-energy state $\ket{\psi_i}$ is determined by the coupling operator $A$ and the corresponding value of the filter function. The coupling operators can often be chosen to be simple operators, such as single Pauli operators or Majorana operators.

The implementation of the jump operator $K$ does not require diagonalizing the Hamiltonian $H$. By expressing the frequency-domain filter $\hat{f}(\omega)$ as the Fourier transform of a time-domain filter function
$f(s) = \frac{1}{2\pi} \int_{\RR} \hat{f}(\omega) e^{-i\omega s} \mathrm{d}\omega$,
the jump operator can be written as
\begin{equation}\label{eqn:jump_time}
K = \int_{-\infty}^{\infty} f(s) e^{iHs} A e^{-iHs} \mathrm{d}s.
\end{equation}
The function $f(s)$ is approximately supported on an interval of size $\Or(\Delta^{-1})$. This allows for efficient truncation and discretization, enabling an efficient quantum implementation of $K$ via the linear combination of unitaries (LCU) method~\cite{ChildsWiebe2012}. The resulting Lindblad dynamics takes the form
\begin{equation}\label{eq:lindblad_ground}
    \frac{\ud}{\ud t} \rho = -i[H, \rho] + K \rho K^\dagger - \frac{1}{2} \{K^\dagger K, \rho\},
\end{equation}
where we have chosen the coherent term $G = H$ to coincide with the system Hamiltonian. For simplicity, we present the case of a single jump operator here and in the following subsections. Multiple jump operators, which typically arising from multiple coupling operators $A$, can be incorporated straightforwardly via linearity.
 It is worth noting that the Lindblad dynamics using a jump operator of the form \cref{eqn:jump_time} can be efficiently implemented, using merely one ancillary qubit~\cite{DingChenLin2024}. 
There are, however, significant challenges analyzing the mixing time of \cref{eq:lindblad_ground}
for pure state preparation. 
Most existing analysis techniques rely on the stationary state $\sigma$ having full rank (see \cref{eqn:gibbs_state}). Consequently, these methods become inapplicable when $\sigma$ is a pure state.

Recent progress has significantly improved the \REV{rigorous theoretical} understanding of the mixing time for dissipative ground state preparation in physically relevant models~\cite{ZhanDingHuhnEtAl2025}.  When the Hamiltonian is quadratic in Majorana operators and the jump operators are linear in Majorana operators, the Lindblad dynamics is called \emph{quasi-free}~\cite{Prosen2008,BarthelZhang2022}. In such settings, the mixing time can be explicitly bounded in terms of the spectral gap of an associated non-Hermitian Hamiltonian. For example, this framework yields an $\Or(n^3)$ mixing time (i.e., fast mixing) for preparing the ground state of a translationally invariant 1D transverse-field Ising model (TFIM) chain, even when dissipation is applied only near the boundaries~\cite{ZhanDingHuhnEtAl2025}. This provides justification for the dilute measurement-induced cooling protocol~\cite{langbehn2024dilute}.

Another direction addresses non-integrable but weakly interacting spin or fermionic systems, where the Hamiltonian takes the form $H = H_0 + \varepsilon H_1$ with $H_0$ gapped and non-interacting. In this regime, it has been shown that the mixing time can be $\log(n)$ (i.e., rapid mixing), provided the perturbation strength $\varepsilon$ is sufficiently small and independent of the system size. The main technical tool is the analysis of the Heisenberg-picture evolution of observables, controlled using the oscillator norm~\cite{rouze2024optimal}. These techniques offer a promising avenue for analyzing more general classes of dissipative dynamics.

\subsection{Local versus quasi-local jump operators for preparing pure states}\label{sec:quasilocal}

A fundamental question in dissipative state preparation is: given a target quantum many-body state $\sigma$, what conditions must the jump operators satisfy so that $\sigma$ is a fixed point of the dynamics? For pure state preparation $\sigma = \ketbra{\psi}$, there are a number of conditions that jump operators should satisfy~\cite{KrausBuchlerDiehlEtAl2008,TicozziViola2012,TicozziViola2014}. Essentially, the target state must be annihilated by each jump operator: \begin{equation}\label{eqn:annihilation_cond}
K_j \ket{\psi} = 0, \quad \forall j. 
\end{equation}
\cref{eqn:annihilation_cond} can severely restrict the type of pure states (e.g., ground states) that can be prepared using \emph{local} dissipative protocols.

To make the concept of local operators concrete, consider a spin system defined on a $D$-dimensional lattice $\Lambda$, and the total number of lattice sites is $n$. An operator $O$ is called \emph{$k$-local} if it acts nontrivially on at most $k$ sites. For example, the one-dimensional transverse field Ising model (TFIM)
\begin{equation}\label{eqn:TFIM_1D}
H = -g \sum_{i=1}^n Z_i - J \sum_{i=1}^{n-1} X_i X_{i+1}
\end{equation}
is a sum of $2$-local terms (note that $1$-local operators are also $2$-local by definition). A jump operator such as $V = X_1 + Y_2$ is a sum of two $1$-local operators but may also be referred to as $2$-local when treated as a single operator acting jointly on qubits $1$ and $2$.

Now assume each $K_j$ is a $k$-local operator, and define a Hermitian operator $\mathsf{H} = \sum_j K_j^{\dagger} K_j$, which is a sum of $k$-local terms. The condition~\eqref{eqn:annihilation_cond} then implies that $\ket{\psi}$ is the simultaneous ground state of all local terms. Hence, $\mathsf{H}$ is a local and frustration-free Hamiltonian. This class includes many useful states, such as stabilizer states and graph states. However, many quantum many-body states (such as most ground states of \emph{ab initio} quantum chemistry Hamiltonians) are \emph{not} ground states of local frustration-free Hamiltonians. This imposes a fundamental limitation on the expressive power of dissipative state preparation when using strictly local jump operators.

It is worth noting that in some experimental settings, the term \emph{quasi-local} may refer to operators acting on a small number of qubits that are nonetheless difficult to implement in practice~\cite{KrausBuchlerDiehlEtAl2008}, distinguishing them from more readily realizable $1$-local or $2$-local operators.
Here we will adopt the convention commonly used in the mathematics and computer science literature, where ``quasi-local'' is used to describe operators that act on geometrically extended regions but whose strength decays rapidly with distance~\cite{NachtergaeleSimsYoung2019,Hastings2019}. As we will see, this  definition allows us to go beyond the limitations of strictly local jump operators and greatly expands the scope of dissipative state preparation.

Again considering a $D$-dimensional lattice $\Lambda$, we now define locality with respect to geometric distance (e.g., Manhattan distance) between lattice sites. For any site $j \in \Lambda$, let $\mathcal{B}_j(r)$ denote the set of sites within distance at most $r - 1$ of $j$; that is, $\mathcal{B}_j(0) = \{j\}$, and $\mathcal{B}_j(1)$ includes $j$ and its nearest neighbors, and so on.  An operator $O_j$ is said to be \emph{$r$-geometrically-local} (or just $r$-local, when the context is clear) around site $j$ if it is supported entirely on $\mathcal{B}_j(r)$. For example, the TFIM Hamiltonian in \cref{eqn:TFIM_1D} is also a sum of $2$-geometrically-local terms.
Now suppose that an operator $O_j$ can be written as a sum $O_j = \sum_{r \in\NN} O_{r,j}$, where each $O_{r,j}$ is supported on $\mathcal{B}_j(r)$ and satisfies
\begin{equation}\label{eqn:quasi_local}
\|O_{r,j}\| \leq C \exp(-\mu r),
\end{equation}
for some constants $C, \mu > 0$ independent of $r$. Then $O_j$ is called a quasi-local operator around site $j$. For example, $O_j=\sum_{k} e^{-\mu \opr{dist}(k,j)} X_k$ is a quasi-local operator in the spin basis around site $j$. The concept of locality and quasi-locality can also be directly generalized to fermionic systems, without being affected by the Jordan--Wigner strings. A key feature of quasi-local operators is that they can be approximated by truncating to a finite support, since their contributions decay exponentially with distance. This makes them amenable to numerical simulation.

In the ground state preparation protocol discussed in \cref{sec:ground_state}, if $A$ is a local operator and $H$ is a sum of local terms, then the Heisenberg-evolved operator $e^{iHs} A e^{-iHs}$ remains approximately supported within a light cone of radius proportional to $\abs{s}$, as dictated by the Lieb-Robinson bound~\cite{LiebRobinson1972,NachtergaeleSims2006,HastingsKoma2006}. Combined with the rapid decay of the function $f(s)$ at large $\abs{s}$, this ensures that the resulting operator $K$ is quasi-local. These quasi-local jump operators are constructed to satisfy the annihilation condition \eqref{eqn:annihilation_cond}, yet they enable dissipative  preparation of states that are not ground states of local frustration-free Hamiltonians~\cite{ZhanDingHuhnEtAl2025}. As a result, dissipative protocols can be a powerful framework for preparing a large class of physically relevant many-body quantum states.

\subsection{Gibbs state preparation}\label{eqn:gibbs_state}

Dissipative protocols can also be used to prepare mixed states, such as the Gibbs state in \cref{eqn:gibbs}.
In recent years, there has been substantial progress in designing
efficient \REV{Monte-Carlo style} algorithms for Gibbs sampling~\cite{TemmeOsborneVollbrechtEtAl2011,moussa2019low,MozgunovLidar2020,ChenBrandao2021,shtanko2021preparing,
RallWangWocjan2023,ChenKastoryanoBrandaoEtAl2023,
ChenKastoryanoGilyen2023,DingLiLin2025,GilyenChenDoriguelloEtAl2024,
JiangIrani2024,moussa2025quantum}.

A central algorithmic theme is the appropriate generalization of the \emph{detailed balance condition} from classical to quantum systems. Classical rejection sampling algorithm, which consists of a proposal step followed by an acceptance or rejection step, leads to a generator $L$ that satisfies the detailed balance condition $L_{ij} \pi_j = L_{ji} \pi_i$, where $\pi$ is the stationary distribution. This guarantees that $\pi$ is a fixed point of the Markov chain.

However, due to the non-commutativity of operators in quantum mechanics, there is no unique definition of
quantum DBC~\cite{TemmeKastoryanoRuskaiEtAl2010,CarlenMaas2017,CarlenMaas2020}.
The most widely studied form of quantum DBC is in the sense of Gelfand--Naimark--Segal (GNS). The seminal result by Alicki~\cite{Alicki1976} characterizes the class of Lindbladians satisfying the GNS DBC, which turns out to have the same dissipative part as that of the Davies generator. An immediate consequence of the GNS DBC is that we need to be able to distinguish the energy levels of $H$ to infinite precision, which cannot be achieved in general, except for some special systems such as commuting Hamiltonians. 
A less restrictive alternative is provided by the Kubo--Martin--Schwinger (KMS) detailed balance condition. However, constructing a Gibbs sampler that exactly satisfies the KMS condition has remained elusive until recently.

Ref.~\cite{ChenKastoryanoGilyen2023} introduced the first efficiently simulatable quantum Gibbs sampler that exactly satisfies the KMS DBC. For a coupling operator $A$ similar to that used in ground state preparation, the key construction is a family of frequency-dependent jump operators defined as
\begin{equation}\label{eqn:freq_jump}
  K(\omega) := \int_{-\infty}^{\infty} f(t) e^{-i\omega t} e^{iHt} A e^{-iHt} \ud t, \quad \omega \in \RR,
\end{equation}
where the filter function $f(t)$ is chosen to be a certain Gaussian function. These jump operators give rise to a Lindbladian of the form
\begin{equation}\label{eqn:lindblad_ckg}
  \mc{L}(\rho) = -i[G,\rho] + \int_{-\infty}^{\infty} \gamma(\omega) \left( K(\omega) \rho K(\omega)^\dagger - \frac{1}{2} \left\{ K(\omega)^\dagger K(\omega), \rho \right\} \right) \ud \omega,
\end{equation}
where $\gamma(\omega): \RR \to [0,1]$ is the transition weight function. A key insight from~\cite{ChenKastoryanoGilyen2023} is that for the Lindbladian to satisfy the KMS DBC, the coherent term $G$ cannot vanish and is in fact uniquely determined by the jump operators $K(\omega)$. With this choice of $G$, one can rigorously show that $\mc{L}(\sigma) = 0$, i.e., the Gibbs state $\sigma$ is a fixed point. Moreover, due to the Gaussian decay of $f(t)$, each $K(\omega)$ is quasi-local and can be efficiently implemented.

What is the most general form of Lindbladians that satisfy the KMS DBC? This question was answered by~\cite{DingLiLin2025}, which shows that it is sufficient to consider a Lindbladian of the form
\begin{equation}\label{eqn:lindblad_dll}
  \mc{L}(\rho) = -i[G, \rho] + K \rho K^\dagger - \frac{1}{2} \left\{ K^\dagger K, \rho \right\},
\end{equation}
where the jump operator $K$ takes the form of \cref{eqn:jump_time}, but with a filter function $f(s)$ designed for thermal state preparation rather than ground state cooling. The coherent term $G$ is again uniquely determined by $K$. Compared to \cref{eqn:lindblad_ckg}, this formulation allows for a single (or few) jump operators, offering significant simplification and flexibility in quantum implementation. As in previous settings, multiple jump operators can be incorporated via linearity. Ref.~\cite{DingLiLin2025} also provides the most general characterization of Lindbladians satisfying the KMS condition when $K$ is of the form \cref{eqn:jump_time}.
Furthermore, Ref.~\cite{GilyenChenDoriguelloEtAl2024} proposes an efficient \emph{discrete-time} approach to quantum Gibbs sampling based on quantum channels that obey a variant of detailed balance. These methods offer a complementary perspective to the continuous-time Lindblad framework. In addition to the Lindblad dynamics framework, quantum Gibbs samplers can also be implemented in a manner more akin to the Monte Carlo framework~\cite{TemmeOsborneVollbrechtEtAl2011,JiangIrani2024}. These approaches offer a promising alternative route for algorithmic design and will not be discussed in this work.

In order to analyze the mixing time of quantum Gibbs samplers, a common strategy  is to examine the spectral gap or the log-Sobolev inequality associated with the Liouvillian. When the quantum detailed balance condition (DBC) is satisfied, the Lindbladian can be mapped via a similarity transformation to a Hermitian operator, enabling the application of \REV{many-body spectral gap techniques~\cite{RouzeFrancaAlhambra2025,tong2024fast,smid2025polynomial}. Since the appearance of Ref.~\cite{ChenKastoryanoGilyen2023}, there has been a surge of rigorous results} concerning the convergence and complexity of quantum Gibbs samplers, particularly in high-temperature and perturbative regimes. We refer readers to~\cite{TemmeKastoryanoRuskaiEtAl2010,KastoryanoTemme2013,BardetCapelGaoEtAl2023,RouzeFrancaAlhambra2025,DingLiLinZhang2024,KochanowskiAlhambraCapelEtAl2025,rouze2024optimal,gamarnik2024slow,tong2024fast,smid2025polynomial,FangLuTong2025}. It is also worth noting that unlike ground state preparation, there is a significant gap between the thermal state preparation of commuting Hamiltonians and frustration-free Hamiltonians. In particular, techniques for quantifying and efficiently preparing thermal states of frustration-free but non-commuting Hamiltonians are still very preliminary due to the lack a common eigenbasis.

The computational complexity of Gibbs sampling is an active area of investigation. For instance, preparing classical Gibbs states at low temperatures is already \NP-hard in the worst case~\cite{Barahona1982,Sly2010}, suggesting that quantum computers are unlikely to efficiently solve such instances in general. At the same time, Ref.~\cite{BakshiLiuMoitraEtAl2024} shows that at high constant temperatures, the Gibbs state of certain $k$-local Hamiltonians can be expressed as a linear combination of tensor products of stabilizer states, and can be prepared via very simple quantum circuits with polynomial-time classical preprocessing.
On the other hand, recent works~\cite{RouzeFrancaAlhambra2025,BergamaschiChenLiu2024,rajakumar2024gibbs} demonstrate that quantum Gibbs samplers can efficiently solve certain sampling tasks believed to be classically intractable. These results suggest that, despite worst-case hardness barriers, quantum approaches may offer a computational advantage in specific regimes or for structured families of instances.

Finally, we note that dissipative algorithms for Gibbs sampling are not only central to quantum simulation, but also serve as a conceptual and algorithmic bridge to classical sampling~\cite{LengDingChenEtAl2025} as well as classical optimization tasks~\cite{ChenLuWangEtAl2025,leng2023quantum}. The interplay between quantum and classical approaches to Gibbs sampling presents a fertile ground for future algorithmic developments.

\subsection{Excited state preparation}\label{sec:excited_state}

In what follows, we propose two further settings, which can be viewed as adaptations of the ground state preparation techniques developed in~\cite{DingChenLin2024}.

In many applications, the goal is to prepare not only the ground state, but also excited states of a quantum Hamiltonian. Excited state preparation is important for simulating molecular absorption spectra and studying non-equilibrium quantum dynamics, and has many applications such as photosynthesis, solar energy harvesting, vision processes, fluorescence microscopy, and designing light-activated molecular devices and sensors.  This can be a  promising target for demonstrating quantum advantage, as  classical  algorithms for excited state computation are far more limited compared to ground state calculations.

There are several strategies for dissipative preparation of excited states. The most natural approach applies when the Hamiltonian possesses a symmetry and the desired excited state lies in a different symmetry sector than the ground state. In this case, by carefully choosing the coupling operator to respect the symmetry of the target excited state, the ground state becomes a ``dark'' state, and the problem can effectively reduce to a ground state preparation within the symmetry-restricted subspace. This is a highly effective and low-cost strategy when applicable, and has been widely adopted, for example, in quantum Monte Carlo setups for targeting low-energy excited states~\cite{FoulkesHoodNeeds1999,ShiZhang2013}.

When the symmetry-based approach is not applicable, but the energy $\mu$ of the target excited state is approximately known, an alternative is to consider the squared Hamiltonian $\wt{H} = (H - \mu I)^2$. In this case, the desired excited state becomes the ground state of $\wt{H}$, and we can apply ground state dissipative preparation techniques to $\wt{H}$. The corresponding jump operator takes the form
\begin{equation}
K = \int_{-\infty}^{\infty} f(s) e^{i(H - \mu I)^2 s} A e^{-i(H - \mu I)^2 s} \mathrm{d}s.
\end{equation}
 While a direct Trotter-based Hamiltonian simulation would require explicitly forming $(H - \mu I)^2$, leading to a proliferation of interaction terms, more advanced techniques such as QSVT treat this as an eigenvalue transformation. Assuming block-encoding access to $H - \mu I$, the function $g(x) = e^{i x^2} = \cos(x^2) + i \sin(x^2)$ can be implemented efficiently via QSVT using its even and odd components~\cite{GilyenSuLowEtAl2019}, or even more directly using a single QSVT circuit~\cite{DongWhaleyLin2022}. However, this comes at the cost of increasing the simulation complexity from $\Or(\norm{H}/\Delta)$ to $\Or(\norm{H}^2/\Delta^2)$.

The $\Or(\norm{H}^2)$ cost may be mitigated in various ways. We propose one possibility that avoids squaring the Hamiltonian:
\begin{equation}
K = P_\mu(H) \int_{-\infty}^{\infty} f(s) e^{iHs} A e^{-iHs} \mathrm{d}s.
\end{equation}
Here $P_\mu(H)$ is a spectral projection onto the subspace with eigenvalues $\lambda \ge \mu$, and $f(s)$ is chosen to be the same as that used in \cref{eqn:jump_time}. When the target excited state energy is the smallest eigenvalue above $\mu$, e.g., $\lambda\ge \mu + \Delta$ for some $\Delta>0$, this projection can be efficiently implemented using QSVT, or alternatively using Hamiltonian simulation through a Fourier representation with a filter function $g_\mu(s)$ as
\begin{equation}
P_\mu(H) = \int g_\mu(s) e^{iHs} \mathrm{d}s.
\end{equation}
In the eigenbasis of $H$, the jump operator becomes
\begin{equation}
K = \sum_{i,j;\,\lambda_i \ge \mu+\Delta} \hat{f}(\lambda_i - \lambda_j) \ket{\psi_i} \bra{\psi_i} A \ket{\psi_j} \bra{\psi_j}.
\end{equation}
This form ensures that only eigenstates with energy above $\mu$ are allowed as targets in the energy-lowering process. 

At first glance, this construction appears to introduce many stationary states: in addition to the desired excited state just above $\mu$, all eigenstates with $\lambda < \mu$ are left untouched and could remain fixed points of the dynamics. However, this issue can be resolved by applying the projection $P_\mu(H)$ to the initial state. Once the initial population in the low-energy subspace is eliminated, these states become dark throughout the evolution and do not affect convergence. \REV{The overall cost of implementing the jump operator can remain $\Or(\norm{H}/\Delta)$, neglecting precision overhead.}
We also expect that certain proof strategies developed for ground state preparations may be useful for excited state preparation as well.

\subsection{Ground singular vector preparation and eigenvector preparation of general matrices}

The method described above for ground state preparation can also be generalized to prepare the right singular vector of a general matrix $T \in \CC^{m \times n}$ associated with its smallest singular value, which we refer to as the
``ground singular vector''. The corresponding left singular vector can be similarly prepared by replacing $T$ with $T^\dag$.

This generalization leverages the fact that the ground singular vector of $T$ is the ground state of the Hermitian positive semidefinite matrix $H = T^\dag T$. Accordingly, we propose the following form for the jump operator:
\begin{equation}\label{eqn:jump_groundsingular}
K = \int_{-\infty}^{\infty} f(s) e^{i T^\dag T s} A e^{-i T^\dag T s} \mathrm{d}s.
\end{equation}
To implement this jump operator efficiently, we assume block-encoding access to $T$, i.e., the ability to construct a unitary $U_T$ such that a properly normalized matrix $T/\alpha$ is embedded as a subblock of $U_T$. Necessarily, $\alpha \ge \norm{T}$, where $\norm{T}$ denotes the operator norm of $T$.

In this construction, the Hamiltonian simulation of $e^{i T^\dag T s}$ can be viewed as a singular value transformation of $T/\alpha$:
\begin{equation}
e^{i (T^\dag / \alpha)(T / \alpha) \alpha^2 s} = g^{(\mathrm{SV})}(T / \alpha),
\end{equation}
where $g^{(\mathrm{SV})}$ denotes the singular value transformation of the function $g(x) = e^{i x^2 \alpha^2 s} = \cos(x^2 \alpha^2 s) + i \sin(x^2 \alpha^2 s)$. This transformation can be implemented efficiently using QSVT, with the cost scaling approximately linearly with the normalized simulation time $\alpha^2 s$. It is also very interesting to design more efficient simulation algorithms whose cost scales as $\Or(\alpha)$ instead of $\Or(\alpha^2)$.

As an application, we can prepare eigenstates of general (non-normal) matrices. This setting arises in problems such as computing resonance states in unbounded domains, such as the Hoyle state of the $^{12}\text{C}$ nucleus~\cite{EpelbaumKrebsLeeEtAl2011,EpelbaumKrebsLaehdeEtAl2012} and electron-attached states of $\mathrm{N}_2$ and $\mathrm{CO}_2$~\cite{JagauBravayaKrylov2017}, evaluating metastability in open quantum systems~\cite{MacieszczakGutaLesanovskyEtAl2016}, and other non-Hermitian quantum systems~\cite{KawabataShiozakiUedaEtAl2019,BergholtzBudichKunst2021}. For any matrix $A \in \CC^{N \times N}$, an eigenpair $A \ket{v} = \lambda \ket{v}$ can be reformulated as a ground singular vector problem for the matrix $\Xi = A - \lambda I$. That is, $\ket{v}$ is the right singular vector of $\Xi$ corresponding to the smallest singular value. We may also need to search over a grid of complex $\lambda$ values to identify approximate eigenpairs and prepare the corresponding eigenvectors.

\section{Opportunities towards quantum advantage and experimental realization}\label{sec:experiment}

\subsection{Potential quantum advantage from dissipative state preparation}
Dissipative state preparation is emerging as a powerful tool for quantum simulation, particularly for preparing complex quantum states that are challenging to access through purely unitary evolution. Ground state preparation in quantum chemistry problems and condensed matter systems  has been widely recognized as a potential route toward achieving quantum advantage. In this case, when variational, adiabatic, or phase estimation methods face challenges due to small energy gaps or difficult initial state preparation, dissipative protocols may offer a more robust alternative by directly driving the system into the desired ground state. 

The quantum preparation of Gibbs states has received significant attention due to its broad applicability. For instance, preparing the ground state of the Fermi-Hubbard model is a key step toward understanding high-temperature superconductivity, while quantum Gibbs sampling algorithms may enable the exploration of finite-temperature phase diagrams. More generally, Gibbs state preparation plays a central role in studying thermodynamic properties of quantum systems, solving optimization problems in quantum machine learning, and enabling quantum-enhanced semidefinite programming and combinatorial optimization.

Excited-state preparation has received comparatively less attention than ground-state and Gibbs-state preparation in quantum algorithms. However, classical methods for excited states are often more limited in scope and scalability. For instance, equation-of-motion coupled cluster (EOM-CC) methods can achieve high accuracy for small molecules but become computationally prohibitive for larger systems due to steep polynomial scaling with the system size, and exponential scaling with respect to the level of the coupled cluster theory.
Tensor network methods may struggle with highly excited states that exhibit volume-law entanglement entropy, requiring exponentially large bond dimensions to accurately represent such states. 

For example, the primary photochemical step in vision involves retinal isomerization via a conical intersection between ground- and excited-state potential energy surfaces~\cite{PolliAltoeWeingartEtAl2010}.
EOM-CC methods up to quadruples can be very accurate for treating small conjugated molecules and clusters, but are difficult to scale to larger systems while achieving sufficient accuracy. In systems such as transition metal complexes and strongly correlated materials, spin and magnetic excitations and strong electron-electron interactions present additional difficulties due to the multi-reference nature of the states.  Moreover, high-energy excitations in correlated materials can lead to cascade processes, including electron scattering and phonon coupling, which require simulation methods that naturally incorporate strong correlations and dynamical evolution.

Resonant state preparation extends these ideas by targeting eigenstates of non-Hermitian Hamiltonians with complex energies $E = E_r - i\Gamma/2$, where the imaginary part $\Gamma$ encodes the decay width. Resonant states arise in many areas of quantum chemistry, such as in the description of radical anions~\cite{JagauBravayaKrylov2017}. Other notable examples include the Hoyle state in $^{12}$C, a long-lived resonance states essential for stellar nucleosynthesis~\cite{EpelbaumKrebsLeeEtAl2011,EpelbaumKrebsLaehdeEtAl2012}, and Feshbach resonances in ultracold atomic gases, which appear as metastable molecular states embedded in a continuum~\cite{ChinGrimmJulienneEtAl2010}.
Classically, the treatment of resonant states can involve a combination of challenges: a suitable one-particle basis, the inherently non-Hermitian formulation, and the inclusion of electron correlation effects.

In these applications, dissipative protocols can bypass the need for carefully engineered initial guesses, which are often even more difficult to construct than those required for ground states.

Another potential direction is the dissipative preparation of bosonic states. Although bosonic systems are formally infinite-dimensional even for finite particle numbers, their spectral properties can differ significantly from those of fermionic systems. Dissipative approaches offer practical paths for error correction in such settings~\cite{LieuLiuGorshkov2024}, and enable algorithmic co-design that integrates error correction with quantum algorithms for bosonic state preparation~\cite{JaegerSchmitMorigiEtAl2022}.

\subsection{Experimental realization}

While the Lindblad framework provides a theoretical foundation for dissipative quantum algorithms, their implementation on quantum hardware poses significant challenges. As discussed in \cref{sec:lindblad_simulate}, the Lindblad dynamics described in this article can, in principle, be simulated efficiently on fault-tolerant quantum computers. \REV{However, direct simulation of Lindblad dynamics using algorithms described in~\cite{CleveWang2017,LiWang2023,DingLiLin2024}} are likely  beyond the reach of near-term and early fault-tolerant devices. It is worth noting that by implementing quantum channels incoherently, the cost of simulating Lindblad dynamics may be significantly reduced, bringing us one step closer to applications on early fault-tolerant quantum computers~\cite{kato2024exponentially,yu2024exponentially}.

Recent experiments on Google's digital quantum processor demonstrated a physics-inspired discrete-time cooling algorithm implemented via a Floquet quantum circuit coupled to an engineered dissipative bath~\cite{MiMichailidisShabaniEtAl2024}. Whether this scheme is sufficient for preparing low-energy states regime relevant for e.g., quantum chemistry applications remains to be tested. 
\REV{What is the connection between the system-bath interaction-based cooling algorithms in~\cite{MiMichailidisShabaniEtAl2024} and the Lindblad dynamics discussed in this article? These two approaches are, in fact, closely related. As discussed in \cref{sec:lindblad_algorithm}, Lindblad dynamics with a Davies-type generator arises from a system-bath interaction Hamiltonian through a sequence of approximations, along with the assumption of ultra-weak coupling. However, ultra-weak coupling implies very long simulation times, which becomes impractical for large systems because the Bohr frequency differences decrease exponentially with system size.}

\REV{Recent empirical and theoretical work strongly suggests that such an ultra-weak coupling regime is likely not necessary for a system-bath interaction Hamiltonian to be well approximated by Lindblad dynamics~\cite{MozgunovLidar2020,langbehn2024dilute,LloydMichailidisMiEtAl2025,lloyd2025quantumthermal,langbehn2025universal,hahn2025provably,scandi2025thermal}. This can be interpreted from two complementary perspectives. On one hand, rigorous bounds on the mixing time of Lindblad dynamics offer theoretical insight into the efficiency of quantum state preparation via system-bath interactions. Indeed, continuous-time Lindblad evolutions may be more amenable to mixing time analysis, and similar behavior also occurs in classical Gibbs sampling~\cite{ChewiErdogduLiEtAl2024,LengDingChenEtAl2025}. On the other hand, the system-bath interaction Hamiltonian itself can be viewed as a physically implementable algorithm for simulating Lindblad dynamics. The interplay between theoretical analysis and physical implementation is likely to yield fruitful developments in the coming years. }

\REV{Inspired by how adiabatic quantum computing motivated the quantum approximate optimization algorithm (QAOA)\cite{FarhiGoldstoneGutmann2014}, another  direction for bringing Lindblad dynamics simulation closer to near-term hardware is the development of dissipative variational circuits. For example, we may parameterize the quantum channel, defined by $U_D$ in \cref{fig:sketch_coherent_dissipative}(b), in order to accelerate cooling. Some initial proposals have appeared recently~\cite{ilin2024dissipative,hahn2025efficient}, and many more remain to be explored.}

These considerations are relevant across hardware platforms, including superconducting qubits, trapped ions, and neutral atoms. Algorithmic co-design tailored to the capabilities of each platform presents a rich opportunity for bridging theory and experiment in the pursuit of quantum advantage.

\section{Outlook}\label{sec:outlook}

Dissipative state preparation is, in many ways, both an old and a new topic. While early developments were largely empirical or tailored to specific structured systems, recent theoretical advances are transforming dissipative techniques into broadly applicable tools for quantum simulation. In particular, the emergence of new algorithmic frameworks in the past few years has enabled more systematic and rigorous design of dissipative protocols across a wide range of settings.

It is important to acknowledge that dissipative algorithms, like all quantum algorithms, are constrained by fundamental complexity-theoretic limitations in the worst case. However, the  performance of dissipative algorithms can vary dramatically between worst-case and physically relevant instances. \REV{This opens up new possibilities for achieving practical quantum advantage, and requires insights at the intersection of computational physics, computer science, and mathematics.}

The potential of achieving quantum advantage via ground state preparation problems have been extensively studied in the literature, and resource estimates for these protocols are actively being refined. Gibbs state preparation has also attracted growing theoretical and experimental interest in recent years.  In the latter part of this essay, we have proposed two comparatively less-explored directions, excited state preparation and resonance state preparation, as promising frontiers for dissipative quantum algorithms. These problems arise naturally in many areas of physics, chemistry and materials science, yet admit few efficient classical or quantum solutions. As such, they may provide alternative paths for demonstrating practical quantum advantage.

The theory and implementation of dissipative state preparation remain rapidly evolving. We expect continued progress in algorithm design, experimental realization, and complexity analysis to deepen our understanding and expand the scope of systems where dissipation can be used as a resource. This makes the coming years an exciting time for both theory and practice in quantum simulation.

\subsection*{Acknowledgments}\label{ssec:acknowledgment}
 This work is partially supported by the Simons Investigator Award in Mathematics.
  We thank Garnet Chan, Zhiyan Ding, Jiaqi Leng, Yu Tong, Lexing Ying and Ruizhe Zhang for valuable suggestions.

\bibliographystyle{abbrvnat}
\bibliography{ref_cite}

\begin{thebibliography}{124}
\providecommand{\natexlab}[1]{#1}
\providecommand{\url}[1]{\texttt{#1}}
\expandafter\ifx\csname urlstyle\endcsname\relax
  \providecommand{\doi}[1]{doi: #1}\else
  \providecommand{\doi}{doi: \begingroup \urlstyle{rm}\Url}\fi

\bibitem[Aharonov and Naveh(2002)]{AharonovNaveh2002}
D.~Aharonov and T.~Naveh.
\newblock Quantum {NP}--a survey.
\newblock \emph{arXiv preprint quant-ph/0210077}, 2002.

\bibitem[Alicki(1976)]{Alicki1976}
R.~Alicki.
\newblock On the detailed balance condition for non-hamiltonian systems.
\newblock \emph{Reports on Mathematical Physics}, 10:\penalty0 249--258, 1976.

\bibitem[An et~al.(2023)An, Childs, and Lin]{an2023quantum}
D.~An, A.~M. Childs, and L.~Lin.
\newblock Quantum algorithm for linear non-unitary dynamics with near-optimal
  dependence on all parameters.
\newblock \emph{arXiv:2312.03916}, 2023.

\bibitem[Bakshi et~al.(2024)Bakshi, Liu, Moitra, and
  Tang]{BakshiLiuMoitraEtAl2024}
A.~Bakshi, A.~Liu, A.~Moitra, and E.~Tang.
\newblock High-temperature gibbs states are unentangled and efficiently
  preparable.
\newblock In \emph{2024 IEEE 65th Annual Symposium on Foundations of Computer
  Science (FOCS)}, pages 1027--1036, 2024.

\bibitem[Barahona(1982)]{Barahona1982}
F.~Barahona.
\newblock On the computational complexity of ising spin glass models.
\newblock \emph{J. Phys. A: Math. Gen.}, 15:\penalty0 3241, 1982.

\bibitem[Bardet et~al.(2023)Bardet, Capel, Gao, Lucia, P{\'e}rez-Garc{\'\i}a,
  and Rouz{\'e}]{BardetCapelGaoEtAl2023}
I.~Bardet, {\'A}.~Capel, L.~Gao, A.~Lucia, D.~P{\'e}rez-Garc{\'\i}a, and
  C.~Rouz{\'e}.
\newblock Rapid thermalization of spin chain commuting hamiltonians.
\newblock \emph{Phys. Rev. Lett.}, 130:\penalty0 060401, 2023.

\bibitem[Barnes and Warren(2000)]{BarnesWarren2000}
J.~P. Barnes and W.~S. Warren.
\newblock Automatic quantum error correction.
\newblock \emph{Phys. Rev. Lett.}, 85:\penalty0 856, 2000.

\bibitem[Barthel and Zhang(2022)]{BarthelZhang2022}
T.~Barthel and Y.~Zhang.
\newblock Solving quasi-free and quadratic lindblad master equations for open
  fermionic and bosonic systems.
\newblock \emph{J. Stat. Mech: Theory Exp.}, 2022:\penalty0 113101, 2022.

\bibitem[Bergamaschi et~al.(2024)Bergamaschi, Chen, and
  Liu]{BergamaschiChenLiu2024}
T.~Bergamaschi, C.-F. Chen, and Y.~Liu.
\newblock Quantum computational advantage with constant-temperature {G}ibbs
  sampling.
\newblock In \emph{2024 IEEE 65th Annual Symposium on Foundations of Computer
  Science (FOCS)}, pages 1063--1085, 2024.

\bibitem[Bergholtz et~al.(2021)Bergholtz, Budich, and
  Kunst]{BergholtzBudichKunst2021}
E.~J. Bergholtz, J.~C. Budich, and F.~K. Kunst.
\newblock Exceptional topology of non-hermitian systems.
\newblock \emph{Rev. Mod. Phys.}, 93:\penalty0 015005, 2021.

\bibitem[Breuer and Petruccione(2002)]{BreuerPetruccione2002}
H.-P. Breuer and F.~Petruccione.
\newblock \emph{The theory of open quantum systems}.
\newblock OUP Oxford, 2002.

\bibitem[Breuer et~al.(2016)Breuer, Laine, Piilo, and
  Vacchini]{BreuerLainePiiloEtAl2016}
H.-P. Breuer, E.-M. Laine, J.~Piilo, and B.~Vacchini.
\newblock Colloquium: Non-markovian dynamics in open quantum systems.
\newblock \emph{Rev. Mod. Phys.}, 88:\penalty0 021002, 2016.

\bibitem[Carlen and Maas(2017)]{CarlenMaas2017}
E.~A. Carlen and J.~Maas.
\newblock Gradient flow and entropy inequalities for quantum markov semigroups
  with detailed balance.
\newblock \emph{J. Funct. Anal.}, 273:\penalty0 1810--1869, 2017.

\bibitem[Carlen and Maas(2020)]{CarlenMaas2020}
E.~A. Carlen and J.~Maas.
\newblock Non-commutative calculus, optimal transport and functional
  inequalities in dissipative quantum systems.
\newblock \emph{J. Stat. Phys.}, 178:\penalty0 319--378, 2020.

\bibitem[Chan(2024)]{Chan2024}
G.~K.-L. Chan.
\newblock Spiers memorial lecture: Quantum chemistry, classical heuristics, and
  quantum advantage.
\newblock \emph{Faraday Discussions}, 254:\penalty0 11--52, 2024.

\bibitem[Chen and Brand{\~a}o(2021)]{ChenBrandao2021}
C.-F. Chen and F.~G. S.~L. Brand{\~a}o.
\newblock Fast thermalization from the eigenstate thermalization hypothesis.
\newblock \emph{arXiv preprint arXiv:2112.07646}, 2021.

\bibitem[Chen et~al.(2023{\natexlab{a}})Chen, Kastoryano, Brand{\~a}o, and
  Gily{\'e}n]{ChenKastoryanoBrandaoEtAl2023}
C.-F. Chen, M.~J. Kastoryano, F.~G. Brand{\~a}o, and A.~Gily{\'e}n.
\newblock Quantum thermal state preparation.
\newblock \emph{arXiv preprint arXiv:2303.18224}, 2023{\natexlab{a}}.

\bibitem[Chen et~al.(2023{\natexlab{b}})Chen, Kastoryano, and
  Gily{\'e}n]{ChenKastoryanoGilyen2023}
C.-F. Chen, M.~J. Kastoryano, and A.~Gily{\'e}n.
\newblock An efficient and exact noncommutative quantum gibbs sampler.
\newblock \emph{arXiv preprint arXiv:2311.09207}, 2023{\natexlab{b}}.

\bibitem[Chen et~al.(2024)Chen, Huang, Preskill, and
  Zhou]{ChenHuangPreskillEtAl2024}
C.-F. Chen, H.-Y. Huang, J.~Preskill, and L.~Zhou.
\newblock Local minima in quantum systems.
\newblock In \emph{Proceedings of the 56th Annual ACM Symposium on Theory of
  Computing}, pages 1323--1330, 2024.

\bibitem[Chen et~al.(2025)Chen, Lu, Wang, Liu, and Li]{ChenLuWangEtAl2025}
Z.~Chen, Y.~Lu, H.~Wang, Y.~Liu, and T.~Li.
\newblock Quantum {L}angevin dynamics for optimization.
\newblock \emph{Commun. Math. Phys.}, 406:\penalty0 52, 2025.

\bibitem[Chewi et~al.(2024)Chewi, Erdogdu, Li, Shen, and
  Zhang]{ChewiErdogduLiEtAl2024}
S.~Chewi, M.~A. Erdogdu, M.~Li, R.~Shen, and M.~S. Zhang.
\newblock {Analysis of Langevin Monte Carlo from Poincar\'e to log-Sobolev}.
\newblock \emph{Found. Comput. Math.}, pages 1--51, 2024.

\bibitem[Childs and Wiebe(2012)]{ChildsWiebe2012}
A.~M. Childs and N.~Wiebe.
\newblock Hamiltonian simulation using linear combinations of unitary
  operations.
\newblock \emph{Quantum Information and Computation}, 12:\penalty0 901--924,
  2012.

\bibitem[Chin et~al.(2010)Chin, Grimm, Julienne, and
  Tiesinga]{ChinGrimmJulienneEtAl2010}
C.~Chin, R.~Grimm, P.~Julienne, and E.~Tiesinga.
\newblock Feshbach resonances in ultracold gases.
\newblock \emph{Rev. Mod. Phys.}, 82:\penalty0 1225--1286, 2010.

\bibitem[Chowdhury and Somma(2017)]{ChowdhurySomma2017}
A.~N. Chowdhury and R.~D. Somma.
\newblock Quantum algorithms for gibbs sampling and hitting-time estimation.
\newblock \emph{Quantum Inf. Comput.}, 17:\penalty0 41--64, 2017.

\bibitem[Cleve and Wang(2017)]{CleveWang2017}
R.~Cleve and C.~Wang.
\newblock Efficient quantum algorithms for simulating {L}indblad evolution.
\newblock In \emph{ICALP 2017}, 2017.

\bibitem[Cubitt(2023)]{Cubitt2023}
T.~S. Cubitt.
\newblock Dissipative ground state preparation and the dissipative quantum
  eigensolver.
\newblock \emph{arXiv:2303.11962}, 2023.

\bibitem[Cubitt et~al.(2015)Cubitt, Lucia, Michalakis, and
  Perez-Garcia]{CubittLuciaMichalakisEtAl2015}
T.~S. Cubitt, A.~Lucia, S.~Michalakis, and D.~Perez-Garcia.
\newblock Stability of local quantum dissipative systems.
\newblock \emph{Commun. Math. Phys.}, 337:\penalty0 1275--1315, 2015.

\bibitem[Davies(1974)]{Davies1974}
E.~B. Davies.
\newblock Markovian master equations.
\newblock \emph{Commun. Math. Phys.}, 39:\penalty0 91--110, 1974.

\bibitem[Davies(1976)]{Davies1976}
E.~B. Davies.
\newblock \emph{Quantum theory of open systems}.
\newblock Academic Press, 1976.

\bibitem[{De Vega} and Alonso(2017)]{DeVegaAlonso2017}
I.~{De Vega} and D.~Alonso.
\newblock Dynamics of non-{M}arkovian open quantum systems.
\newblock \emph{Rev. Mod. Phys.}, 89:\penalty0 1--58, 2017.

\bibitem[Ding et~al.(2024{\natexlab{a}})Ding, Chen, and Lin]{DingChenLin2024}
Z.~Ding, C.-F. Chen, and L.~Lin.
\newblock Single-ancilla ground state preparation via {L}indbladians.
\newblock \emph{Phys. Rev. Research}, 6:\penalty0 033147, 2024{\natexlab{a}}.

\bibitem[Ding et~al.(2024{\natexlab{b}})Ding, Li, Lin, and
  Zhang]{DingLiLinZhang2024}
Z.~Ding, B.~Li, L.~Lin, and R.~Zhang.
\newblock Polynomial-time preparation of low-temperature {G}ibbs states for {2D
  Toric Code}.
\newblock \emph{arXiv:2410.01206}, 2024{\natexlab{b}}.

\bibitem[Ding et~al.(2024{\natexlab{c}})Ding, Li, and Lin]{DingLiLin2024}
Z.~Ding, X.~Li, and L.~Lin.
\newblock Simulating open quantum systems using {H}amiltonian simulations.
\newblock \emph{PRX Quantum}, 5:\penalty0 020332, 2024{\natexlab{c}}.

\bibitem[Ding et~al.(2025)Ding, Li, and Lin]{DingLiLin2025}
Z.~Ding, B.~Li, and L.~Lin.
\newblock Efficient quantum gibbs samplers with kubo--martin--schwinger
  detailed balance condition.
\newblock \emph{Commun. Math. Phys.}, 406:\penalty0 67, 2025.

\bibitem[Dong et~al.(2022{\natexlab{a}})Dong, Lin, and Tong]{DongLinTong2022}
Y.~Dong, L.~Lin, and Y.~Tong.
\newblock Ground-state preparation and energy estimation on early
  fault-tolerant quantum computers via quantum eigenvalue transformation of
  unitary matrices.
\newblock \emph{PRX Quantum}, 3:\penalty0 040305, 2022{\natexlab{a}}.

\bibitem[Dong et~al.(2022{\natexlab{b}})Dong, Whaley, and
  Lin]{DongWhaleyLin2022}
Y.~Dong, K.~B. Whaley, and L.~Lin.
\newblock A quantum {H}amiltonian simulation benchmark.
\newblock \emph{npj Quantum Inf.}, 8:\penalty0 131, 2022{\natexlab{b}}.

\bibitem[Eder et~al.(2025)Eder, Fin{\v{z}}gar, Braun, and
  Mendl]{EderFinzgarBraunEtAl2025}
P.~J. Eder, J.~R. Fin{\v{z}}gar, S.~Braun, and C.~B. Mendl.
\newblock Quantum dissipative search via {L}indbladians.
\newblock \emph{Phys. Rev. A}, 111:\penalty0 042430, 2025.

\bibitem[Epelbaum et~al.(2011)Epelbaum, Krebs, Lee, and
  Mei{\ss}ner]{EpelbaumKrebsLeeEtAl2011}
E.~Epelbaum, H.~Krebs, D.~Lee, and U.-G. Mei{\ss}ner.
\newblock Ab initio calculation of the hoyle state.
\newblock \emph{Phys. Rev. Lett.}, 106:\penalty0 192501, 2011.

\bibitem[Epelbaum et~al.(2012)Epelbaum, Krebs, L{\"a}hde, Lee, and
  Mei{\ss}ner]{EpelbaumKrebsLaehdeEtAl2012}
E.~Epelbaum, H.~Krebs, T.~A. L{\"a}hde, D.~Lee, and U.-G. Mei{\ss}ner.
\newblock Structure and rotations of the hoyle state.
\newblock \emph{Phys. Rev. Lett.}, 109:\penalty0 252501, 2012.

\bibitem[Fang et~al.(2025)Fang, Lu, and Tong]{FangLuTong2025}
D.~Fang, J.~Lu, and Y.~Tong.
\newblock Mixing time of open quantum systems via hypocoercivity.
\newblock \emph{Phys. Rev. Lett.}, 134:\penalty0 140405, 2025.

\bibitem[Farhi et~al.(2014)Farhi, Goldstone, and
  Gutmann]{FarhiGoldstoneGutmann2014}
E.~Farhi, J.~Goldstone, and S.~Gutmann.
\newblock A quantum approximate optimization algorithm.
\newblock \emph{arXiv:1411.4028}, 2014.

\bibitem[Foss-Feig et~al.(2023)Foss-Feig, Tikku, Lu, Mayer, Iqbal, Gatterman,
  Gerber, Gilmore, Gresh, Hankin, et~al.]{foss2023experimental}
M.~Foss-Feig, A.~Tikku, T.-C. Lu, K.~Mayer, M.~Iqbal, T.~M. Gatterman, J.~A.
  Gerber, K.~Gilmore, D.~Gresh, A.~Hankin, et~al.
\newblock Experimental demonstration of the advantage of adaptive quantum
  circuits.
\newblock \emph{arXiv preprint arXiv:2302.03029}, 2023.

\bibitem[Foulkes et~al.(1999)Foulkes, Hood, and Needs]{FoulkesHoodNeeds1999}
W.~Foulkes, R.~Q. Hood, and R.~Needs.
\newblock Symmetry constraints and variational principles in diffusion quantum
  monte carlo calculations of excited-state energies.
\newblock \emph{Phys. Rev. B}, 60:\penalty0 4558, 1999.

\bibitem[Frenkel and Smit(2002)]{FrenkelSmit2002}
D.~Frenkel and B.~Smit.
\newblock \emph{Understanding Molecular Simulation: From Algorithms to
  Applications}.
\newblock Academic Press, 2002.

\bibitem[Gamarnik et~al.(2024)Gamarnik, Kiani, and Zlokapa]{gamarnik2024slow}
D.~Gamarnik, B.~T. Kiani, and A.~Zlokapa.
\newblock Slow mixing of quantum {Gibbs} samplers.
\newblock \emph{arXiv preprint arXiv:2411.04300}, 2024.

\bibitem[Ge et~al.(2019)Ge, Tura, and Cirac]{GeTuraCirac2019}
Y.~Ge, J.~Tura, and J.~I. Cirac.
\newblock Faster ground state preparation and high-precision ground energy
  estimation with fewer qubits.
\newblock \emph{J. Math. Phys.}, 60:\penalty0 022202, 2019.

\bibitem[Gily{\'e}n et~al.(2019)Gily{\'e}n, Su, Low, and
  Wiebe]{GilyenSuLowEtAl2019}
A.~Gily{\'e}n, Y.~Su, G.~H. Low, and N.~Wiebe.
\newblock Quantum singular value transformation and beyond: exponential
  improvements for quantum matrix arithmetics.
\newblock In \emph{Proceedings of the 51st Annual ACM SIGACT Symposium on
  Theory of Computing}, pages 193--204, 2019.

\bibitem[Gily\'en et~al.(2024)Gily\'en, Chen, Doriguello, and
  Kastoryano]{GilyenChenDoriguelloEtAl2024}
A.~Gily\'en, C.-F. Chen, J.~F. Doriguello, and M.~J. Kastoryano.
\newblock Quantum generalizations of {G}lauber and {M}etropolis dynamics.
\newblock \emph{arXiv:2405.20322}, 2024.

\bibitem[Gorini et~al.(1976)Gorini, Kossakowski, and
  Sudarshan]{GoriniKossakowskiSudarshan1976}
V.~Gorini, A.~Kossakowski, and E.~C.~G. Sudarshan.
\newblock Completely positive dynamical semigroups of $n$-level systems.
\newblock \emph{J. Math. Phys.}, 17:\penalty0 821--825, 1976.

\bibitem[Hahn et~al.(2025{\natexlab{a}})Hahn, Parameswaran, and
  Placke]{hahn2025provably}
D.~Hahn, S.~A. Parameswaran, and B.~Placke.
\newblock Provably efficient quantum thermal state preparation via local
  driving.
\newblock \emph{arXiv:2505.22816}, 2025{\natexlab{a}}.

\bibitem[Hahn et~al.(2025{\natexlab{b}})Hahn, Sweke, Deshpande, and
  Shtanko]{hahn2025efficient}
D.~Hahn, R.~Sweke, A.~Deshpande, and O.~Shtanko.
\newblock Efficient quantum {Gibbs} sampling with local circuits.
\newblock \emph{arXiv:2506.04321}, 2025{\natexlab{b}}.

\bibitem[Hastings(2019)]{Hastings2019}
M.~B. Hastings.
\newblock The stability of free fermi hamiltonians.
\newblock \emph{J. Math. Phys.}, 60:\penalty0 042201, 2019.

\bibitem[Hastings and Koma(2006)]{HastingsKoma2006}
M.~B. Hastings and T.~Koma.
\newblock Spectral gap and exponential decay of correlations.
\newblock \emph{Commun. Math. Phys.}, 265:\penalty0 781--804, 2006.

\bibitem[Ilin and Arad(2024)]{ilin2024dissipative}
Y.~Ilin and I.~Arad.
\newblock Dissipative variational quantum algorithms for {G}ibbs state
  preparation.
\newblock \emph{IEEE Trans. Quantum Eng.}, 2024.

\bibitem[Jagau et~al.(2017)Jagau, Bravaya, and Krylov]{JagauBravayaKrylov2017}
T.~C. Jagau, K.~B. Bravaya, and A.~I. Krylov.
\newblock Extending quantum chemistry of bound states to electronic resonances.
\newblock \emph{Annu. Rev. Phys. Chem.}, 68:\penalty0 525--553, 2017.

\bibitem[J{\"a}ger et~al.(2022)J{\"a}ger, Schmit, Morigi, Holland, and
  Betzholz]{JaegerSchmitMorigiEtAl2022}
S.~B. J{\"a}ger, T.~Schmit, G.~Morigi, M.~J. Holland, and R.~Betzholz.
\newblock Lindblad master equations for quantum systems coupled to dissipative
  bosonic modes.
\newblock \emph{Phys. Rev. Lett.}, 129:\penalty0 063601, 2022.

\bibitem[Jiang and Irani(2024)]{JiangIrani2024}
J.~Jiang and S.~Irani.
\newblock Quantum metropolis sampling via weak measurement.
\newblock \emph{arXiv preprint arXiv:2406.16023}, 2024.

\bibitem[Kashyap et~al.(2025)Kashyap, Styliaris, Mouradian, Cirac, and
  Trivedi]{KashyapStyliarisMouradianEtAl2025}
V.~Kashyap, G.~Styliaris, S.~Mouradian, J.~I. Cirac, and R.~Trivedi.
\newblock Accuracy guarantees and quantum advantage in analog open quantum
  simulation with and without noise.
\newblock \emph{Phys. Rev. X}, 15:\penalty0 021017, 2025.

\bibitem[Kastoryano and Temme(2013)]{KastoryanoTemme2013}
M.~J. Kastoryano and K.~Temme.
\newblock Quantum logarithmic sobolev inequalities and rapid mixing.
\newblock \emph{J. Math. Phys.}, 54:\penalty0 1--34, 2013.

\bibitem[Kato et~al.(2024)Kato, Wada, Ito, and Yamamoto]{kato2024exponentially}
J.~Kato, K.~Wada, K.~Ito, and N.~Yamamoto.
\newblock Exponentially accurate open quantum simulation via randomized
  dissipation with minimal ancilla.
\newblock \emph{arXiv preprint arXiv:2412.19453}, 2024.

\bibitem[Kawabata et~al.(2019)Kawabata, Shiozaki, Ueda, and
  Sato]{KawabataShiozakiUedaEtAl2019}
K.~Kawabata, K.~Shiozaki, M.~Ueda, and M.~Sato.
\newblock Symmetry and topology in non-hermitian physics.
\newblock \emph{Phys. Rev. X}, 9:\penalty0 041015, 2019.

\bibitem[Kempe et~al.(2006)Kempe, Kitaev, and Regev]{KempeKitaevRegev2006}
J.~Kempe, A.~Kitaev, and O.~Regev.
\newblock The complexity of the local {Hamiltonian} problem.
\newblock \emph{SIAM J. Comput.}, 35:\penalty0 1070--1097, 2006.

\bibitem[Kitaev et~al.(2002)Kitaev, Shen, and Vyalyi]{KitaevShenVyalyi2002}
A.~Y. Kitaev, A.~Shen, and M.~N. Vyalyi.
\newblock \emph{Classical and quantum computation}.
\newblock American Mathematical Soc., 2002.

\bibitem[Kliesch et~al.(2011)Kliesch, Barthel, Gogolin, Kastoryano, and
  Eisert]{KlieschBarthelGogolinEtAl2011}
M.~Kliesch, T.~Barthel, C.~Gogolin, M.~J. Kastoryano, and J.~Eisert.
\newblock Dissipative quantum {C}hurch-{T}uring theorem.
\newblock \emph{Phys. Rev. Lett.}, 107, 2011.

\bibitem[Kochanowski et~al.(2025)Kochanowski, Alhambra, Capel, and
  Rouz{\'e}]{KochanowskiAlhambraCapelEtAl2025}
J.~Kochanowski, A.~M. Alhambra, A.~Capel, and C.~Rouz{\'e}.
\newblock Rapid thermalization of dissipative many-body dynamics of commuting
  hamiltonians.
\newblock \emph{Commun. Math. Phys.}, 406:\penalty0 176, 2025.

\bibitem[Kraus et~al.(2008)Kraus, B\"uchler, Diehl, Kantian, Micheli, and
  Zoller]{KrausBuchlerDiehlEtAl2008}
B.~Kraus, H.~P. B\"uchler, S.~Diehl, A.~Kantian, A.~Micheli, and P.~Zoller.
\newblock Preparation of entangled states by quantum markov processes.
\newblock \emph{Phys. Rev. A}, 78:\penalty0 042307, 2008.

\bibitem[Lambert et~al.(2024)Lambert, Cirio, Lin, Menczel, Liang, and
  Nori]{LambertCirioLinEtAl2024}
N.~Lambert, M.~Cirio, J.-D. Lin, P.~Menczel, P.~Liang, and F.~Nori.
\newblock Fixing detailed balance in ancilla-based dissipative state
  engineering.
\newblock \emph{Phys. Rev. Research}, 6:\penalty0 043229, 2024.

\bibitem[Landi et~al.(2022)Landi, Poletti, and
  Schaller]{LandiPolettiSchaller2022}
G.~T. Landi, D.~Poletti, and G.~Schaller.
\newblock Nonequilibrium boundary-driven quantum systems: Models, methods, and
  properties.
\newblock \emph{Rev. Mod. Phys.}, 94:\penalty0 045006, 2022.

\bibitem[Langbehn et~al.(2024)Langbehn, Snizhko, Gornyi, Morigi, Gefen, and
  Koch]{langbehn2024dilute}
J.~Langbehn, K.~Snizhko, I.~Gornyi, G.~Morigi, Y.~Gefen, and C.~P. Koch.
\newblock Dilute measurement-induced cooling into many-body ground states.
\newblock \emph{PRX Quantum}, 5:\penalty0 030301, 2024.

\bibitem[Langbehn et~al.(2025)Langbehn, Mouloudakis, King, Menu, Gornyi,
  Morigi, Gefen, and Koch]{langbehn2025universal}
J.~Langbehn, G.~Mouloudakis, E.~King, R.~Menu, I.~Gornyi, G.~Morigi, Y.~Gefen,
  and C.~P. Koch.
\newblock Universal cooling of quantum systems via randomized measurements.
\newblock \emph{arXiv:2506.11964}, 2025.

\bibitem[Lee et~al.(2023)Lee, Lee, Zhai, Tong, Dalzell, Kumar, Helms, Gray,
  Cui, Liu, et~al.]{LeeLeeZhaiEtAl2023}
S.~Lee, J.~Lee, H.~Zhai, Y.~Tong, A.~M. Dalzell, A.~Kumar, P.~Helms, J.~Gray,
  Z.-H. Cui, W.~Liu, et~al.
\newblock Evaluating the evidence for exponential quantum advantage in
  ground-state quantum chemistry.
\newblock \emph{Nature Comm.}, 14:\penalty0 1952, 2023.

\bibitem[Leghtas et~al.(2013)Leghtas, Kirchmair, Vlastakis, Schoelkopf,
  Devoret, and Mirrahimi]{LeghtasKirchmairVlastakisEtAl2013}
Z.~Leghtas, G.~Kirchmair, B.~Vlastakis, R.~J. Schoelkopf, M.~H. Devoret, and
  M.~Mirrahimi.
\newblock Hardware-efficient autonomous quantum memory protection.
\newblock \emph{Phys. Rev. Lett.}, 111:\penalty0 120501, 2013.

\bibitem[Leng et~al.(2023)Leng, Hickman, Li, and Wu]{leng2023quantum}
J.~Leng, E.~Hickman, J.~Li, and X.~Wu.
\newblock Quantum {H}amiltonian descent.
\newblock \emph{arXiv preprint arXiv:2303.01471}, 2023.

\bibitem[Leng et~al.(2025)Leng, Ding, Chen, and Lin]{LengDingChenEtAl2025}
J.~Leng, Z.~Ding, Z.~Chen, and L.~Lin.
\newblock Operator-level quantum acceleration of non-logconcave sampling.
\newblock \emph{arXiv:2505.05301}, 2025.

\bibitem[Levin and Peres(2017)]{LevinPeres2017}
D.~A. Levin and Y.~Peres.
\newblock \emph{Markov chains and mixing times}, volume 107.
\newblock American Mathematical Soc., 2017.

\bibitem[Li et~al.(2024)Li, Zhan, and Lin]{li2024dissipative}
H.-E. Li, Y.~Zhan, and L.~Lin.
\newblock Dissipative ground state preparation in ab initio electronic
  structure theory.
\newblock \emph{arXiv preprint arXiv:2411.01470}, 2024.

\bibitem[Li and Wang(2023)]{LiWang2023}
X.~Li and C.~Wang.
\newblock Simulating markovian open quantum systems using higher-order series
  expansion.
\newblock In \emph{50th International Colloquium on Automata, Languages, and
  Programming (ICALP 2023)}, volume 261, pages 87:1--87:20, 2023.

\bibitem[Lidar(2019)]{Lidar2019}
D.~A. Lidar.
\newblock Lecture notes on the theory of open quantum systems, 2019.

\bibitem[Lieb and Robinson(1972)]{LiebRobinson1972}
E.~H. Lieb and D.~W. Robinson.
\newblock The finite group velocity of quantum spin systems.
\newblock \emph{Commun. Math. Phys.}, 28:\penalty0 251--257, 1972.

\bibitem[Lieu et~al.(2024)Lieu, Liu, and Gorshkov]{LieuLiuGorshkov2024}
S.~Lieu, Y.~J. Liu, and A.~V. Gorshkov.
\newblock Candidate for a passively protected quantum memory in two dimensions.
\newblock \emph{Phys. Rev. Lett.}, 133:\penalty0 30601, 2024.

\bibitem[Lin and Tong(2020)]{LinTong2020}
L.~Lin and Y.~Tong.
\newblock Optimal quantum eigenstate filtering with application to solving
  quantum linear systems.
\newblock \emph{Quantum}, 4:\penalty0 361, 2020.

\bibitem[Lin and Tong(2022)]{LinTong2022}
L.~Lin and Y.~Tong.
\newblock Heisenberg-limited ground state energy estimation for early
  fault-tolerant quantum computers.
\newblock \emph{PRX Quantum}, 3:\penalty0 010318, 2022.

\bibitem[Lindblad(1976)]{Lindblad1976}
G.~Lindblad.
\newblock On the generators of quantum dynamical semigroups.
\newblock \emph{Commun. Math. Phys.}, 48:\penalty0 119--130, 1976.

\bibitem[Lloyd and Abanin(2025)]{lloyd2025quantumthermal}
J.~Lloyd and D.~A. Abanin.
\newblock Quantum thermal state preparation for near-term quantum processors.
\newblock \emph{arXiv:2506.21318}, 2025.

\bibitem[Lloyd et~al.(2025)Lloyd, Michailidis, Mi, Smelyanskiy, and
  Abanin]{LloydMichailidisMiEtAl2025}
J.~Lloyd, A.~A. Michailidis, X.~Mi, V.~Smelyanskiy, and D.~A. Abanin.
\newblock Quasiparticle cooling algorithms for quantum many-body state
  preparation.
\newblock \emph{PRX Quantum}, 6:\penalty0 010361, 2025.

\bibitem[Lu et~al.(2022)Lu, Lessa, Kim, and Hsieh]{LuLessaKimEtAl2022}
T.-C. Lu, L.~A. Lessa, I.~H. Kim, and T.~H. Hsieh.
\newblock Measurement as a shortcut to long-range entangled quantum matter.
\newblock \emph{PRX Quantum}, 3:\penalty0 040337, 2022.

\bibitem[Macieszczak et~al.(2016)Macieszczak, Gu{\c{t}}{\u{a}}, Lesanovsky, and
  Garrahan]{MacieszczakGutaLesanovskyEtAl2016}
K.~Macieszczak, M.~Gu{\c{t}}{\u{a}}, I.~Lesanovsky, and J.~P. Garrahan.
\newblock Towards a theory of metastability in open quantum dynamics.
\newblock \emph{Phys. Rev. Lett.}, 116:\penalty0 240404, 2016.

\bibitem[Mi et~al.(2024)Mi, Michailidis, Shabani, Miao, Klimov, Lloyd,
  Rosenberg, Acharya, Aleiner, Andersen, Ansmann, Arute, Arya, Asfaw, Atalaya,
  Bardin, Bengtsson, Bortoli, Bourassa, Bovaird, Brill, Broughton, Buckley,
  Buell, Burger, Burkett, Bushnell, Chen, Chiaro, Chik, Chou, Cogan, Collins,
  Conner, Courtney, Crook, Curtin, Dau, Debroy, {Del Toro Barba}, Demura, {Di
  Paolo}, Drozdov, Dunsworth, Erickson, Faoro, Farhi, Fatemi, Ferreira, Burgos,
  Forati, Fowler, Foxen, Genois, Giang, Gidney, Gilboa, Giustina, Gosula,
  Gross, Habegger, Hamilton, Hansen, Harrigan, Harrington, Heu, Hoffmann, Hong,
  Huang, Huff, Huggins, Ioffe, Isakov, Iveland, Jeffrey, Jiang, Jones, Juhas,
  Kafri, Kechedzhi, Khattar, Khezri, Kieferov{\'{a}}, Kim, Kitaev, Klots,
  Korotkov, Kostritsa, Kreikebaum, Landhuis, Laptev, Lau, Laws, Lee, Lee,
  Lensky, Lester, Lill, Liu, Locharla, Malone, Martin, McClean, McEwen,
  Mieszala, Montazeri, Morvan, Movassagh, Mruczkiewicz, Neeley, Neill,
  Nersisyan, Newman, Ng, Nguyen, Nguyen, Niu, O'Brien, Opremcak, Petukhov,
  Potter, Pryadko, Quintana, Rocque, Rubin, Saei, Sank, Sankaragomathi,
  Satzinger, Schurkus, Schuster, Shearn, Shorter, Shutty, Shvarts, Skruzny,
  Smith, Somma, Sterling, Strain, Szalay, Torres, Vidal, Villalonga,
  Heidweiller, White, Woo, Xing, Yao, Yeh, Yoo, Young, Zalcman, Zhang, Zhu,
  Zobrist, Neven, Babbush, Bacon, Boixo, Hilton, Lucero, Megrant, Kelly, Chen,
  Roushan, Smelyanskiy, and Abanin]{MiMichailidisShabaniEtAl2024}
X.~Mi, A.~A. Michailidis, S.~Shabani, K.~C. Miao, P.~V. Klimov, J.~Lloyd,
  E.~Rosenberg, R.~Acharya, I.~Aleiner, T.~I. Andersen, M.~Ansmann, F.~Arute,
  K.~Arya, A.~Asfaw, J.~Atalaya, J.~C. Bardin, A.~Bengtsson, G.~Bortoli,
  A.~Bourassa, J.~Bovaird, L.~Brill, M.~Broughton, B.~B. Buckley, D.~A. Buell,
  T.~Burger, B.~Burkett, N.~Bushnell, Z.~Chen, B.~Chiaro, D.~Chik, C.~Chou,
  J.~Cogan, R.~Collins, P.~Conner, W.~Courtney, A.~L. Crook, B.~Curtin, A.~G.
  Dau, D.~M. Debroy, A.~{Del Toro Barba}, S.~Demura, A.~{Di Paolo}, I.~K.
  Drozdov, A.~Dunsworth, C.~Erickson, L.~Faoro, E.~Farhi, R.~Fatemi, V.~S.
  Ferreira, L.~F. Burgos, E.~Forati, A.~G. Fowler, B.~Foxen, {\'{E}}.~Genois,
  W.~Giang, C.~Gidney, D.~Gilboa, M.~Giustina, R.~Gosula, J.~A. Gross,
  S.~Habegger, M.~C. Hamilton, M.~Hansen, M.~P. Harrigan, S.~D. Harrington,
  P.~Heu, M.~R. Hoffmann, S.~Hong, T.~Huang, A.~Huff, W.~J. Huggins, L.~B.
  Ioffe, S.~V. Isakov, J.~Iveland, E.~Jeffrey, Z.~Jiang, C.~Jones, P.~Juhas,
  D.~Kafri, K.~Kechedzhi, T.~Khattar, M.~Khezri, M.~Kieferov{\'{a}}, S.~Kim,
  A.~Kitaev, A.~R. Klots, A.~N. Korotkov, F.~Kostritsa, J.~M. Kreikebaum,
  D.~Landhuis, P.~Laptev, K.-M. Lau, L.~Laws, J.~Lee, K.~W. Lee, Y.~D. Lensky,
  B.~J. Lester, A.~T. Lill, W.~Liu, A.~Locharla, F.~D. Malone, O.~Martin, J.~R.
  McClean, M.~McEwen, A.~Mieszala, S.~Montazeri, A.~Morvan, R.~Movassagh,
  W.~Mruczkiewicz, M.~Neeley, C.~Neill, A.~Nersisyan, M.~Newman, J.~H. Ng,
  A.~Nguyen, M.~Nguyen, M.~Y. Niu, T.~E. O'Brien, A.~Opremcak, A.~Petukhov,
  R.~Potter, L.~P. Pryadko, C.~Quintana, C.~Rocque, N.~C. Rubin, N.~Saei,
  D.~Sank, K.~Sankaragomathi, K.~J. Satzinger, H.~F. Schurkus, C.~Schuster,
  M.~J. Shearn, A.~Shorter, N.~Shutty, V.~Shvarts, J.~Skruzny, W.~C. Smith,
  R.~Somma, G.~Sterling, D.~Strain, M.~Szalay, A.~Torres, G.~Vidal,
  B.~Villalonga, C.~V. Heidweiller, T.~White, B.~W.~K. Woo, C.~Xing, Z.~J. Yao,
  P.~Yeh, J.~Yoo, G.~Young, A.~Zalcman, Y.~Zhang, N.~Zhu, N.~Zobrist, H.~Neven,
  R.~Babbush, D.~Bacon, S.~Boixo, J.~Hilton, E.~Lucero, A.~Megrant, J.~Kelly,
  Y.~Chen, P.~Roushan, V.~Smelyanskiy, and D.~A. Abanin.
\newblock Stable quantum-correlated many-body states through engineered
  dissipation.
\newblock \emph{Science}, 383:\penalty0 1332--1337, 2024.

\bibitem[Motlagh et~al.(2024)Motlagh, Zini, Arrazola, and
  Wiebe]{motlagh2024ground}
D.~Motlagh, M.~S. Zini, J.~M. Arrazola, and N.~Wiebe.
\newblock Ground state preparation via dynamical cooling.
\newblock \emph{arXiv preprint arXiv:2404.05810}, 2024.

\bibitem[Moussa(2019)]{moussa2019low}
J.~E. Moussa.
\newblock Low-depth quantum {M}etropolis algorithm.
\newblock \emph{arXiv preprint arXiv:1903.01451}, 2019.

\bibitem[Moussa(2025)]{moussa2025quantum}
J.~E. Moussa.
\newblock Quantum {M}etropolis-{H}astings algorithm.
\newblock \emph{arXiv preprint arXiv:2503.14970}, 2025.

\bibitem[Mozgunov and Lidar(2020)]{MozgunovLidar2020}
E.~Mozgunov and D.~Lidar.
\newblock Completely positive master equation for arbitrary driving and small
  level spacing.
\newblock \emph{Quantum}, 4:\penalty0 1--62, 2020.

\bibitem[Nachtergaele and Sims(2006)]{NachtergaeleSims2006}
B.~Nachtergaele and R.~Sims.
\newblock Lieb-robinson bounds and the exponential clustering theorem.
\newblock \emph{Commun. Math. Phys.}, 265:\penalty0 119--130, 2006.

\bibitem[Nachtergaele et~al.(2019)Nachtergaele, Sims, and
  Young]{NachtergaeleSimsYoung2019}
B.~Nachtergaele, R.~Sims, and A.~Young.
\newblock Quasi-locality bounds for quantum lattice systems. i. lieb-robinson
  bounds, quasi-local maps, and spectral flow automorphisms.
\newblock \emph{J. Math. Phys.}, 60:\penalty0 061101, 2019.

\bibitem[Nielsen and Chuang(2000)]{NielsenChuang2000}
M.~A. Nielsen and I.~Chuang.
\newblock Quantum computation and quantum information, 2000.

\bibitem[O'Brien et~al.(2019)O'Brien, Tarasinski, and
  Terhal]{OBrienTarasinskiTerhal2019}
T.~E. O'Brien, B.~Tarasinski, and B.~M. Terhal.
\newblock Quantum phase estimation of multiple eigenvalues for small-scale
  (noisy) experiments.
\newblock \emph{New J. Phys.}, 21:\penalty0 023022, 2019.

\bibitem[Polli et~al.(2010)Polli, Alto{\`e}, Weingart, Spillane, Manzoni,
  Brida, Tomasello, Orlandi, Kukura, Mathies,
  et~al.]{PolliAltoeWeingartEtAl2010}
D.~Polli, P.~Alto{\`e}, O.~Weingart, K.~M. Spillane, C.~Manzoni, D.~Brida,
  G.~Tomasello, G.~Orlandi, P.~Kukura, R.~A. Mathies, et~al.
\newblock Conical intersection dynamics of the primary photoisomerization event
  in vision.
\newblock \emph{Nature}, 467:\penalty0 440--443, 2010.

\bibitem[Poulin and Wocjan(2009)]{PoulinWocjan2009}
D.~Poulin and P.~Wocjan.
\newblock Preparing ground states of quantum many-body systems on a quantum
  computer.
\newblock \emph{Phys. Rev. Lett.}, 102:\penalty0 130503, 2009.

\bibitem[Prosen(2008)]{Prosen2008}
T.~Prosen.
\newblock Third quantization: a general method to solve master equations for
  quadratic open fermi systems.
\newblock \emph{New J. Phys.}, 10:\penalty0 043026, 2008.

\bibitem[Rajakumar and Watson(2024)]{rajakumar2024gibbs}
J.~Rajakumar and J.~D. Watson.
\newblock Gibbs {S}ampling gives {Q}uantum {A}dvantage at {C}onstant
  {T}emperatures with ${O} (1) $-{L}ocal {H}amiltonians.
\newblock \emph{arXiv preprint arXiv:2408.01516}, 2024.

\bibitem[Rall et~al.(2023)Rall, Wang, and Wocjan]{RallWangWocjan2023}
P.~Rall, C.~Wang, and P.~Wocjan.
\newblock Thermal state preparation via rounding promises.
\newblock \emph{Quantum}, 7:\penalty0 1132, 2023.

\bibitem[Rouz{\'e} et~al.(2024)Rouz{\'e}, Fran{\c{c}}a, and
  Alhambra]{rouze2024optimal}
C.~Rouz{\'e}, D.~S. Fran{\c{c}}a, and {\'A}.~M. Alhambra.
\newblock Optimal quantum algorithm for {Gibbs} state preparation.
\newblock \emph{arXiv:2411.04885}, 2024.

\bibitem[Rouz{\'e} et~al.(2025)Rouz{\'e}, Fran{\c{c}}a, and
  Alhambra]{RouzeFrancaAlhambra2025}
C.~Rouz{\'e}, D.~S. Fran{\c{c}}a, and {\'A}.~M. Alhambra.
\newblock Efficient thermalization and universal quantum computing with quantum
  {G}ibbs samplers.
\newblock In \emph{Proceedings of the 57th Annual ACM Symposium on Theory of
  Computing}, pages 1488--1495, 2025.

\bibitem[Roy et~al.(2020)Roy, Chalker, Gornyi, and
  Gefen]{RoyChalkerGornyiEtAl2020}
S.~Roy, J.~Chalker, I.~Gornyi, and Y.~Gefen.
\newblock Measurement induced steering of quantum systems.
\newblock \emph{Phys. Rev. Research}, 2:\penalty0 033347, 2020.

\bibitem[Scandi and Alhambra(2025)]{scandi2025thermal}
M.~Scandi and A.~M. Alhambra.
\newblock Thermalization in open many-body systems and {KMS} detailed balance.
\newblock \emph{arXiv:2505.20064}, 2025.

\bibitem[Shi and Zhang(2013)]{ShiZhang2013}
H.~Shi and S.~Zhang.
\newblock Symmetry in auxiliary-field quantum monte carlo calculations.
\newblock \emph{Phys. Rev. B}, 88:\penalty0 125132, 2013.

\bibitem[Shtanko and Movassagh(2021)]{shtanko2021preparing}
O.~Shtanko and R.~Movassagh.
\newblock Preparing thermal states on noiseless and noisy programmable quantum
  processors.
\newblock \emph{arXiv:2112.14688}, 2021.

\bibitem[Sly(2010)]{Sly2010}
A.~Sly.
\newblock Computational transition at the uniqueness threshold.
\newblock In \emph{2010 IEEE 51st Annual Symposium on Foundations of Computer
  Science}, pages 287--296, 2010.

\bibitem[{\v{S}}m{\'\i}d et~al.(2025){\v{S}}m{\'\i}d, Meister, Berta, and
  Bondesan]{smid2025polynomial}
{\v{S}}.~{\v{S}}m{\'\i}d, R.~Meister, M.~Berta, and R.~Bondesan.
\newblock Polynomial time quantum {G}ibbs sampling for {Fermi-Hubbard} model at
  any temperature.
\newblock \emph{arXiv preprint arXiv:2501.01412}, 2025.

\bibitem[Temme et~al.(2010)Temme, Kastoryano, Ruskai, Wolf, and
  Verstraete]{TemmeKastoryanoRuskaiEtAl2010}
K.~Temme, M.~J. Kastoryano, M.~B. Ruskai, M.~M. Wolf, and F.~Verstraete.
\newblock The $\chi^2$-divergence and mixing times of quantum markov processes.
\newblock \emph{J. Math. Phys.}, 51, 2010.

\bibitem[Temme et~al.(2011)Temme, Osborne, Vollbrecht, Poulin, and
  Verstraete]{TemmeOsborneVollbrechtEtAl2011}
K.~Temme, T.~J. Osborne, K.~G. Vollbrecht, D.~Poulin, and F.~Verstraete.
\newblock Quantum {Metropolis} sampling.
\newblock \emph{Nature}, 471:\penalty0 87--90, 2011.

\bibitem[Ticozzi and Viola(2012)]{TicozziViola2012}
F.~Ticozzi and L.~Viola.
\newblock Stabilizing entangled states with quasi-local quantum dynamical
  semigroups.
\newblock \emph{Philosophical Transactions of the Royal Society A:
  Mathematical, Physical and Engineering Sciences}, 370:\penalty0 5259--5269,
  2012.

\bibitem[Ticozzi and Viola(2014)]{TicozziViola2014}
F.~Ticozzi and L.~Viola.
\newblock Steady-state entanglement by engineered quasi-local markovian
  dissipation.
\newblock \emph{Quantum Information and Computation}, 14:\penalty0 265--294,
  2014.

\bibitem[Tong and Zhan(2024)]{tong2024fast}
Y.~Tong and Y.~Zhan.
\newblock Fast mixing of weakly interacting fermionic systems at any
  temperature.
\newblock \emph{arXiv:2501.00443}, 2024.

\bibitem[Trivedi et~al.(2024)Trivedi, {Franco Rubio}, and
  Cirac]{TrivediFrancoRubioCirac2024}
R.~Trivedi, A.~{Franco Rubio}, and J.~I. Cirac.
\newblock Quantum advantage and stability to errors in analogue quantum
  simulators.
\newblock \emph{Nature Commun.}, 15:\penalty0 6507, 2024.

\bibitem[Van~Apeldoorn et~al.(2017)Van~Apeldoorn, Gily{\'e}n, Gribling, and
  de~Wolf]{VanApeldoornGilyenGriblingEtAl2017}
J.~Van~Apeldoorn, A.~Gily{\'e}n, S.~Gribling, and R.~de~Wolf.
\newblock Quantum sdp-solvers: Better upper and lower bounds.
\newblock In \emph{2017 IEEE 58th Annual Symposium on Foundations of Computer
  Science (FOCS)}, pages 403--414, 2017.

\bibitem[Verstraete et~al.(2009)Verstraete, Wolf, and
  Cirac]{VerstraeteWolfCirac2009}
F.~Verstraete, M.~M. Wolf, and I.~Cirac.
\newblock Quantum computation and quantum-state engineering driven by
  dissipation.
\newblock \emph{Nat. Phys.}, 5:\penalty0 633--636, 2009.

\bibitem[Wan et~al.(2022)Wan, Berta, and Campbell]{WanBertaCampbell2022}
K.~Wan, M.~Berta, and E.~T. Campbell.
\newblock Randomized quantum algorithm for statistical phase estimation.
\newblock \emph{Phys. Rev. Lett.}, 129:\penalty0 030503, 2022.

\bibitem[Wang et~al.(2023)Wang, Snizhko, Romito, Gefen, and
  Murch]{WangSnizhkoRomitoEtAl2023}
Y.~Wang, K.~Snizhko, A.~Romito, Y.~Gefen, and K.~Murch.
\newblock Dissipative preparation and stabilization of many-body quantum states
  in a superconducting qutrit array.
\newblock \emph{Phys. Rev. A}, 108:\penalty0 013712, 2023.

\bibitem[Watrous(2018)]{Watrous2018}
J.~Watrous.
\newblock \emph{The theory of quantum information}.
\newblock Cambridge Univ. Pr., 2018.

\bibitem[Weimer et~al.(2021)Weimer, Kshetrimayum, and
  Or{\'u}s]{WeimerKshetrimayumOrus2021}
H.~Weimer, A.~Kshetrimayum, and R.~Or{\'u}s.
\newblock Simulation methods for open quantum many-body systems.
\newblock \emph{Rev. Mod. Phys.}, 93:\penalty0 015008, 2021.

\bibitem[Wolf(2012)]{Wolf2012}
M.~M. Wolf.
\newblock Quantum channels and operations-guided tour, 2012.

\bibitem[Yu et~al.(2024)Yu, Li, Zhao, and Yuan]{yu2024exponentially}
W.~Yu, X.~Li, Q.~Zhao, and X.~Yuan.
\newblock Exponentially reduced circuit depths in lindbladian simulation.
\newblock \emph{arXiv preprint arXiv:2412.21062}, 2024.

\bibitem[Zhan et~al.(2025)Zhan, Ding, Huhn, Gray, Preskill, Chan, and
  Lin]{ZhanDingHuhnEtAl2025}
Y.~Zhan, Z.~Ding, J.~Huhn, J.~Gray, J.~Preskill, G.~K. Chan, and L.~Lin.
\newblock Rapid quantum ground state preparation via dissipative dynamics.
\newblock \emph{arXiv preprint arXiv:2503.15827}, 2025.

\end{thebibliography}

\end{document}